\documentclass[preprint,10pt]{aastex}

\shorttitle{Chromospheric Variability}
\shortauthors{Rachford \& Foight}

\begin{document}

\title{Chromospheric Variability in Early F-type Stars}

\author{Brian L. Rachford\altaffilmark{1}, Dillon R. Foight}
\affil{Department of Physics, Embry-Riddle Aeronautical
University, 3700 Willow Creek Road, Prescott, AZ 86301;
rachf7ac@erau.edu}
\altaffiltext{1}{Visiting Astronomer, Kitt Peak National
Observatory, National Optical Astronomy Observatories, which is
operated by the Association of Universities for Research in
Astronomy Inc., under cooperative agreement with the National
Science Foundation.}

\begin{abstract}
Using precise measurements of the helium D$_3$ line, we have
searched for statistically significant variations in the
strength of chromospheric activity in 13 early F-type stars and
two late F-type stars.  In two early F-type stars, we find
short-term (hours to days) variability based on $\sim$25
observations over the course of a week.  In an additional two
cases we find significant differences between observations taken
years apart, but we can most likely explain this apparent
long-term variation as an artifact of probable short-term
variations.  The evidence suggests that pure rotational
modulation of discrete active regions is not responsible for the
short-term variations in the early F-type stars and that either
a more global process is at work, or we are seeing large number
of small active regions spread across the star.  In contrast,
the two late F-type stars in the sample show strength and/or
wavelength variations that are consistent with ``solar-type''
activity typified by the rotational modulation of active
regions.  Our results suggest that variability does not cause
the wide range in activity levels observed within the early
F-type stars.
\end{abstract}


\setcounter{footnote}{1}
\section{Introduction}
Chromospheric variability on rotational time scales has been
well-demonstrated through a variety of spectral indicators in
late F- through K-type stars, due to discrete active regions
crossing the visible disk
\citep[e.g.,][]{bal83,ayr99,fra00,bia07}.  As in the Sun, these
active regions are thought to be associated with concentrated
magnetic fields.  In addition, long-term cycles (years to
decades) analogous to the 11-year solar cycle have been found in
Sun-like stars \citep{bal95}.

The situation with regards to chromospheric variability in the
early F-type stars is less clear.  The standard activity
indicators become more difficult to observe and there is
uncertainty as to the role of magnetic fields in driving the
activity resulting from the very thin near-surface convective
regions \citep[e.g.,][]{nar96,nef08}.  Nearly all early F-type
stars show at least moderately strong activity, with few if any
analogs to the weak activity of the Sun
\citep{sim91,gar93,rac97} and no apparent dependence on age
\citep{rac00}.  However, within a narrow temperature range or at
a specific age across all early F-type stars, a large activity
range exists.

Numerous UV emission lines, which stand out against the weak UV
continua of these stars, as well as the optical \ion{He}{1}
$\lambda$5876 (D$_3$) absorption line have been successfully
used as chromospheric activity indicators in the early F-type
stars.  These indicators have occasionally been used in attempts
to search for variability, which is one possible explanation for
the activity range in these stars.  \citet{ayr91} used the {\it
IUE} satellite to observe several UV emission lines in
Hertzsprung gap star $\beta$ Cas in an attempt to find
rotational modulation.  In twenty spectra the author measured
three lines, Ly$\alpha$, \ion{C}{2} $\lambda$1335, and
\ion{C}{4} $\lambda$1549.  The observations of the two carbon
lines showed standard deviations 20--50\% larger than the
typical measurement uncertainties while the Ly$\alpha$
observations showed standard deviations more than twice as great
as the typical measurement uncertainties.  However, Ly$\alpha$
measurements are greatly affected by interstellar absorption and
geocoronal emission, and in addition the author indicated that
the measurement uncertainties may have been underestimated in
some cases.

\citet{wol86} observed the D$_3$ line in $\mu$ Vir 12 times, but
attributed the spectrum-to-spectrum variation to inaccuracies in
the removal of telluric line contamination.  \citet{rac98}
reported evidence for D$_3$ variability in relatively
low-precision data obtained for two early F-type open cluster
stars, one in Coma and the other in Praesepe.  Despite very
limited temporal coverage, the data suggested short-term
variability on time scales of less than one day.  Finally,
\citet{ter05} observed the D$_3$ line in a variety of late-type
stars, including several F-type stars.  Teresova reported
evidence for short-term D$_3$ variability in one early F-type
star and possible evidence for long-term variability based on
comparisons with D$_3$ results from other authors.

To further explore chromospheric variability in the early F
stars, we have undertaken an observing program involving
high-precision measurements of D$_3$ in 15 bright stars.  A key
component of this study is the uniform processing and analysis
of the data, particularly important because we must remove
photospheric and telluric lines that contaminate the D$_3$ line.
In addition, we pay careful attention to the uncertainties in
our measurements, crucial for quantifying variability.  Thus, we
describe our procedures in some detail, not only in support of
the present work but also for related future studies using a
broader dataset.

We organize the rest of the paper as follows.  In \S\ 2 we
describe the observations and the reduction to 1-dimensional
spectra.  In \S\ 3 we describe in detail our analysis procedures
that yield D$_3$ line parameters.  In \S\ 4 we give an overview
of our stellar sample.  In \S\ 5 we describe the results of our
search for short-term D$_3$ variability.  In \S\ 6 we describe
the results of our search for long-term variability.  In \S\ 7
we discuss our results and their implications.  Finally, we
summarize the paper in \S\ 8.

\section{Observations and Data Reduction}
All data were obtained with the now defunct 0.9-meter Coud\'{e}
Feed Telescope at Kitt Peak National Observatory, using the
echelle grating, Camera 5, and the F3KB CCD.  We used a 300
$\mu$m slit corresponding to 2.1 arcsec projected on the sky and
the dispersion was 25 m\AA\ pixel$^{-1}$ near the D$_3$ line.
Based on measurements of Th-Ar lamp spectra, we obtained 2.8
pixel resolution (70 m\AA , 3.6 km s$^{-1}$, or
$\lambda/\Delta\lambda$ $\approx$ 85000).  Table 1 gives an
overview of the observing runs during which we obtained data.

We used standard routines from the IRAF\footnote{IRAF is
distributed by the National Optical Astronomy Observatories,
operated by AURA Inc., under cooperative agreement with the
National Science Foundation} ECHELLE package for bias
subtraction, flat-fielding, scattered light removal, reduction
to one-dimensional spectra, continuum normalization, and
wavelength calibration.  For the latter, we obtained several
Th-Ar spectra each night, and by fitting $\sim$30 lines per
echelle order, we obtained $\sim$50 m s$^{-1}$ RMS uncertainty
in the wavelength scale on most nights.

With the large horizontal format of the CCD (3072 pixels), we
cover $\approx$75 \AA\ in each order with about 20\% overlap
across adjacent orders.  To perform spectral fits across a wide
wavelength range, we must normalize each echelle order and
``stitch'' together the adjacent orders.  The IRAF CONTINUUM
package allowed us to interactively fine-tune the continuum
fitting to best exclude low-flux points associated with
photospheric and telluric lines.  Thus, we could use a
relatively high polynomial order without distorting the stellar
lines.  As discussed in \S\ 3.6, the final measurement of the
D$_3$ line involves a small portion of an echelle order which
can be fitted with a low-order continuum.

\section{Data Analysis}

\subsection{Overview}
Precise measurement of the D$_3$ line or any other spectral
line requires that the line be as unaffected as possible by
other spectral features.  For D$_3$, adjacent photospheric lines
as well as telluric water vapor lines can be serious
contaminants.  The latter is particularly problematic because it
is highly variable.  Figure 1 illustrates the problem and
demonstrates our ability to correct for these contaminants.  As
this figure shows, much of the problem lies with the telluric
lines superposed on the D$_3$ line, but it is also important to
remove photospheric lines for proper continuum rectification.

We remove the photospheric lines by fitting model spectra to a
large portion of our observed spectra to determine stellar
rotational velocity and abundances of the relevant chemical
elements, and then using this information to generate a model
spectrum for the immediate D$_3$ region.  We remove the telluric
lines by fitting a model to the observed target spectrum, and
then using that information to properly scale a very high S/N
telluric template spectrum to match the target spectrum.

In both cases we take advantage of the fact that there are
photospheric and telluric lines in other portions of the
spectrum that are stronger than the lines that must be removed
from the D$_3$ vicinity.  Thus, the absolute errors in the
photospheric and telluric fits are minimized when removing the
weaker lines.  Once we have removed the contaminating
photospheric and telluric lines, it is in principle a simple
matter to measure the remaining D$_3$ line.  In the following
sections we describe the entire process in more detail.

\subsection{Synthetic Spectral Fitting}
We synthesized spectra with the SPECTRUM program \citep{gra94},
which requires as inputs a line list and a stellar atmospheric
model.  For the former, we started with the list of solar lines
given by \citet{the90}, supplemented with Robert Kurucz's on-line
lists\footnote{http://cfaku5.harvard.edu/}.  We synthesized each
line individually for both a solar model ($T_{\rm eff}$ = 5777 K
and log $g$ = 4.4377) and a late A star ($T_{\rm eff}$ = 8000 K
and log $g$ = 4.0) to see which lines are important in the
temperature range of our data, deleting lines with equivalent
widths less than 0.1 m\AA\ in both models.  Through an iterative
process we adjusted some of the oscillator strengths to match
our own solar spectrum (the lunar disk observed with the same
equipment as the stellar observations).  Except for a small
percentage of lines in the Kurucz list that were grossly in
error, most differences were small, and we only adjusted lines
which clearly did not match our observed spectrum.  For these
adjustments, we used the solar abundances tabulated by
\citet[][pg. 319, see references within]{gra88}.

We assume that the photospheric lines remain constant in the
stars, so to determine the elemental abundances we optimized the
S/N to provide a ``key'' spectrum for each star.  Target
information will be given in \S\ 4, but for stars with a small
number of observations, we chose the one with the highest S/N
(excluding data from a run with limited wavelength coverage),
and for the stars that were observed several times in a night we
coadded all spectra from a single night to give a very high S/N
key spectrum.  We chose the range 6007--6200 \AA\ to maximize
the number of elements we could fit and minimize the presence of
telluric lines as this latter correction is best determined with
accurate knowledge of the photospheric spectrum.

For the key fits we used the Levenburg-Marquardt non-linear
least squares method \citep[i.e., the ``CURFIT'' algorithm
from][]{bev92}.  The free parameters in the fits are the
zero-point wavelength shift, microturbulent speed, $v$ sin $i$,
and the logarithmic elemental abundances.  In addition to
rotational broadening, we include instrumental broadening (a
Gaussian with FWHM of $\sim$70 m\AA\ based on the Th-Ar lines
used for wavelength calibration), and macroturbulent broadening
(a polynomial approximation to the radial-tangential
macroturbulence function given by \citet[][pg. 1-18]{gra88}
corresponding to 5 km s$^{-1}$ for the F-type stars).  Both
additional broadenings are generally small compared to $v$ sin
$i$ for our stellar sample.

We only perform the key fit for one spectrum (or one summed
spectrum) for a star and use that information to produce a model
spectrum for the D$_3$ area in all spectra.  However, we need an
accurate wavelength shift to properly apply this model spectrum
to each additional spectrum.  Thus, we fit the 5845--5865 \AA\
region, which contains enough lines to provide an accurate shift
and is in the same original echelle order as D$_3$, eliminating
possible differences between the wavelength solutions in
different orders.  However, in this region only calcium, iron,
and barium have strong enough lines for abundance
determinations.  We perform this fit for all spectra using the
downhill simplex method \citep[the ``AMOEBA'' algorithm
from][]{pre00}, which we prefer over ``CURFIT'' when fitting
poorly known parameters.  The information from these fits can
then be used to provide good guesses for the parameters of the
key fit for each star, which includes more elements as
discussed below.

We use the wavelength shift from the 5845--5865 \AA\ region to
derive heliocentric velocities using the IRAF RVCORRECT
procedure.  We report these velocities along with the D$_3$ line
parameters, primarily as a rough indicator of binarity.  We do
not report formal uncertainties for each of these measurements
as the ``AMEOBA'' algorithm does not report them, but they
typically range from about 0.2--0.3 km s$^{-1}$ for the stars
with the smallest $v$ sin $i$ to as much as 1--2 km s$^{-1}$ for
the stars with the largest $v$ sin $i$.  These uncertainties are
far smaller than those for the D$_3$ central wavelengths that we
report and are thus not a significant source of error for these
wavelengths.

In addition to the two fitted spectral regions, we must model the
5880--5904 \AA\ range (also in the same original echelle order
as D$_3$) to properly determine the strength of telluric
absorption, and the 5865--5885 \AA\ region to remove the
contaminating lines from the D$_3$ vicinity.  The net result is
that we end up modeling the entire range from 5845--5904 \AA .

We used 72-level \citet{kur92} stellar models, originally
generated on a temperature grid with spacing 250 K and log $g$
spacing of 0.5 dex.  For our fits, we interpolated the models to
provide 125 K and 0.1 dex spacing, respectively, for temperature
and gravity.  The models assume a depth-independent
microturbulence of 2.0 km s$^{-1}$, a reasonable choice for our
stars as the actual derived microturbulences from the model fits
are in the range 1--4 km s$^{-1}$.

We used the software developed by \citet{nap93} to derive the
effective temperature and surface gravity from uvby$\beta$
photometry taken from the \citet{hau98} catalog.  We then
rounded these values to match our Kurucz model grid and used
that specific model for a particular star.  The \citet{nap93}
code also calculates stellar radii based a relationship between
the Str\"{o}mgrem photometry and surface brightness derived by
\citet{moo84}, and a slightly modified version of the
\citet{bar78} relationship between surface brightness, absolute
magnitude, and radius.  However, we used {\it Hipparcos}
absolute magnitudes instead of those calculated by the
\citet{nap93} code.  The stellar radii in turn allow us to
calculate the projected rotational periods for the stars.

In the key fits, we allowed the abundances of up to nine
elements to vary, including the elements that are most likely to
interfere with D$_3$.  These elements are listed in Table 2
which also gives our derived solar abundances, discussed in \S\
3.4.  For our early F-type stars we could not always reliably
determine abundances for all nine elements, so we only report
values for the five elements most accurately fitted in all 15
stars.  This still includes all species that significantly
interfere with the D$_3$ region in F-type stars.  Weaker lines
from approximately 20 additional elements were modeled using
solar abundances.  The SPECTRUM program calculates model spectra
using the LTE assumption.  In the early F stars, this is a
reasonable approximation for the weak to moderate line strengths
we cover.

\subsection{Verification of the spectral modeling}
We tested our fitting procedures on spectra of the Sun (the
lunar disk) and $\alpha$ CMi (Procyon; F5 IV--V), one of our
D$_3$ program stars.  Since we used the solar spectrum to adjust
the $gf$ factors, we would expect the fit to this spectrum to be
an excellent match to other studies and it is.  Table 2 shows
the results of the solar fit for a spectrum with a S/N $\approx$
500, using a model with $T_{\rm eff}$ = 5770 K and log $g$ =
4.4377.  We found a reasonable microturbulent velocity, $\xi_t$
= 1.04 km s$^{-1}$, and a rotational broadening, $v$ sin $i$ =
2.13 km s$^{-1}$, that closely matches the solar equatorial
rotation speed of 1.98 km s$^{-1}$.

Table 3 lists similar data for a spectrum of Procyon with S/N
$\approx$ 300, using a model with $T_{\rm eff}$ = 6625 K and log
$g$ = 4.1.  We found $\xi_t$ = 1.85 km s$^{-1}$ and $v$ sin $i$
= 5.78 km s$^{-1}$.  These values compare well with other
studies; e.g., $\xi_t$ = 2.1 km s$^{-1}$ by \citet{ste85} from
a curve-of-growth analysis and $\xi_t$ = 1.9 km s$^{-1}$ by
\citet{var99} via a similar procedure to ours.  \citet{fek97}
derived $v$ sin $i$ = 4.9 km s$^{-1}$ and \citet{var99}
found $v$ sin $i$ = 7 km s$^{-1}$.

\citet{var99} used a model with $T_{\rm eff}$ = 6696 K, so to
provide a further comparision to our fit, we also performed a
fit with $T_{\rm eff}$ = 6750 K.  This choice brackets the
\citet{var99} fit with the two most similar models from our grid
and we give the results of both of our fits in Table 3.
Additionally, this provides an estimate of the uncertainties in
our quoted abundances based on the temperature uncertainty, which
\citet{nap93} report as about 2.5\% ($\sim$160--180 K) in the
temperature range covered by our sample.  We see that the
resulting abundance uncertainties resulting from the temperature
calibration are less than 0.1 dex for the five elements reported
for the rest of our sample.

\subsection{Telluric Fit}
Our procedure for removing telluric water vapor lines is
essentially identical to that used in previous papers
\citep{rac98,rac00} so we only give a brief overview here.

\citet{lun91} catalogued the telluric water vapor lines in the
range 5868--5917 \AA , which accounts for virtually all telluric
absorption in this wavelength range.  The strongest telluric
lines within this range are near the sodium D lines, which were
covered by the same echelle order as D$_3$ in our spectra.  With
$v$ sin $i$ and wavelength shifts known from the 5845--5865 \AA\
synthetic spectral fits, we generated an approximate synthetic
photospheric spectrum for the range 5884--5904 \AA\ to flag
pixels that are significantly affected by photospheric lines and
totally exclude them from the telluric model fit.  We then
fitted the telluric model to the remaining pixels based on the
line catalog.  Since the water vapor lines scale together, this
gives one parameter describing the strength of the lines for
that exposure.

We produced a very high S/N template of the telluric spectrum by
observing a rapidly rotating, unreddened hot star (Regulus) and
using the CONTINUUM routine in IRAF to ``flatten'' the few
shallow photospheric lines.  We scaled the template based on the
telluric line strength in each target spectrum derived from the
model fits, and this scaled template provides the telluric
removal.  By using this technique, we avoided issues with minor
imperfections in the modeling of the weak telluric lines near
D$_3$ since we divided through an actual observed spectrum
instead of the model.

\subsection{Final Processing}
After dividing the target spectrum by the photospheric model and
the scaled telluric spectrum, we are left with a spectrum that
should only contain D$_3$ absorption.  As with the telluric
correction, our techniques are similar to those used in previous
work \citep{rac98,rac00}.  One potential problem with the
construction of the original target spectra is that we must use
a relatively high-order polynomial fit to normalize each order
of the spectrum, mostly due to the echelle blaze function.
While this works well for the spectrum as a whole, there is the
risk of small differences from spectrum to spectrum due to the
variability of the telluric lines and putative variability in
the D$_3$ line, which may change the exact pixels that are
included in the continuum fit.  For instance, more pixels may be
excluded in the D$_3$ vicinity if the telluric and D$_3$
absorption are strong.  While we did not see any obvious
problems in the normalized spectra, we wanted to be conservative
in our search for variability.  The original continuum can
accurately be fitted with a third-order polynomial in the narrow
range around D$_3$, thus we applied the photospheric and
telluric corrections to the unnormalized spectra.

The D$_3$ line is typically broadened in excess of the stellar
$v$ sin $i$ \citep{rac00}.  Thus, while the observed D$_3$
profiles are not simply Gaussian nor the $v$ sin $i$ ``bowl''
function, at the level of precision of our spectra a Gaussian
provides a good match to the profiles.  Importantly, this gives
us an analytic fitting function for which it is easy to generate
formal uncertainties in our derived equivalent widths, crucial
for assessing variability.  Thus, we performed a 7-parameter fit
for each spectrum using the CURFIT routine: central wavelength
($\lambda_0$), Gaussian width ($\sigma$), Gaussian depth ($d$),
and a 3rd order continuum.  The equivalent width of a Gaussian
absorption profile in a normalized spectrum is then simply
\begin{equation}
W_{\lambda} = \sqrt{2 \pi} \sigma d
\end{equation}
with the final uncertainty in equivalent width calculated with
standard error propagation techniques based on the formal
uncertainties in $\sigma$ and $d$.

\subsection{Errors}
To assess the existence of variability, we must be certain that
the uncertainties on the individual D$_3$ measurements are
accurately determined.  To verify that the uncertainties
reported by the fitting routine were correct, we performed Monte
Carlo simulations whereby we generated ``perfect'' D$_3$
profiles (plus continuum) that matched the results of fits for
several stellar spectra.  Then we added a large number of random
noise vectors to each profile corresponding to the measured S/N
of the original spectrum.  We then fitted these simulated noisy
profiles in the same way as the actual data.

We found that the simulated profiles produced D$_3$ equivalent
widths with uncertainties and standard deviations within 15\% of
the originally derived values for that spectrum, while the mean
equivalent widths of the simulated lines provided nearly exact
matches to the actual lines.  The results for the central
wavelengths and line widths were similar, and indicate that we
are not strongly over or underestimating the uncertainties on
the individual measurements we report.  However, given the
relatively small sample sizes, a 15\% underestimate of the
uncertainties is sometimes enough to significantly affect the
interpretation of the results and this will be discussed in more
detail in \S\ 5.1.  We mostly attribute the small differences
between the actual spectra and the simulations to slight
inaccuracies in the telluric and photospheric line removals, and
slight differences between the true shape of the observed D$_3$
lines and a Gaussian.

\section{Target Selection}
With the exception of four stars that were extensively observed
throughout a specific observing run, most of the observations
were taken during non-optimal observing conditions when we could
not observe fainter targets; i.e., light to moderate cloudiness
or twilight.  Thus, we were limited to stars with $V$ $\lesssim$
5.  In addition, the weak D$_3$ line becomes much more difficult
to detect with large rotational broadening, so we limited the
sample to $v$ sin $i$ $\lesssim$ 100 km s$^{-1}$.  Our goal was
to obtain continuum S/N ratios around 200--300, yielding D$_3$
equivalent width precision of about 5--10\% for targets with
``normal'' line strengths; this goal was not achieved in all
cases.

Table 4 lists the program stars along with pertinent photometric
and spectroscopic data.  In addition to data from the {\it
Hipparcos Catalog} \citep{esa97} we also quote a more uniform
set of modern spectral types that provide a much better match to
our effective temperatures and surface gravities calculated as
described in \S\ 3.2.  Figure 2 gives the positions of each star
in a color-magnitude diagram of nearby stars observed by {\it
Hipparcos}.  In Table 5, we give the results of our synthetic
spectral fits, including microturbulence, rotational velocity,
rotational period, and elemental abundances.  Our sample can be
broadly described as ranging from one-half solar metallicity to
full solar metallicity, and our [Fe/H] values agree well with
previous measurements.

We had particular reasons for observing two of the four stars
for which we investigated short-term variability.  We observed
18 Boo due to highly discrepant literature values of the D$_3$
equivalent width (40 m\AA\ from \citet{wol86} and 10 $\pm$ 5
m\AA\ from \citet{gar93}.  In addition, as discussed in \S\ 1,
$\mu$ Vir has already been the subject of a search for D$_3$
variability.  Although our observations were made before the
\citet{ter05} study, there are several additional targets in
common between our studies.

We have reported D$_3$ equivalent widths for nine of these stars
in a previous work \citep{rac97}.  We reanalyzed those spectra for
the present study to provide a uniform comparison with the newer
observations.  In particular, our photospheric fitting and
modeling procedure has improved by determining specific
abundances for each element for each star, and we model the
D$_3$ lines as Gaussians in the present work as opposed to a
wavelength-by-wavelength summation.  Our present equivalent
widths are generally slightly smaller than the previously
reported values, which we mostly attribute to better
photospheric line removal, but otherwise the agreement is
reasonable.  We emphasize that it is somewhat difficult to
precisely compare sets of D$_3$ measurements that have not been
processed in exactly the same way due to the telluric and
photospheric line removals.

\section{Short-term Variability}

\subsection{Overall results and statistical methods}
In Tables 6 through 9 we give observing information and
resulting D$_3$ parameters for the 100 spectra of four stars we
used for the investigation of short-term variability, including
a few spectra taken to search for long-term variability which
will be discussed in \S\ 6.2.  The S/N was determined in the
vicinity of the D$_3$ line, but varies somewhat across the
overall spectrum.  We also give heliocentric radial velocities,
which illustrate that we do not see significant
radial velocity variability for these stars.  We have numbered
the spectra in the tables so we can refer to them individually
in further discussion.  In all cases, we have also listed the
observing run for each observation using the same notation as
Table 1 to provide a simpler context for the timing of the
observations than the Julian Dates.

We give the central wavelengths of the D$_3$ lines in the rest
frame of the star, based on the photospheric line fits in the
5845--5865 \AA\ range.  The reported uncertainties do not
explicitly include the wavelength uncertainty of the
photospheric fits or possible run-to-run differences in the
wavelength solutions, but these errors are typically much
smaller than the errors in the D$_3$ fits.  For convenience, we
have converted the Gaussian linewidths into the full width at
half maximum expressed in km s$^{-1}$, which can then be
compared with $v$ sin $i$ for the star.

To explore the existence of short-term variability, we begin by
calculating various statistics on the measurements of equivalent
width, central wavelength, and linewidth.  Table 10 gives a
summary of these values.  For each quantity we have tabulated
the weighted mean, the weighted standard deviation, and the mean
uncertainty of the individual measurements.  We chose the mean
uncertainty instead of the median uncertainty because the mean
was usually slightly larger, giving a slightly more conservative
basis for comparing the standard deviation and measurement
uncertainties.  Still, a comparison between the standard
deviation and typical uncertainties immediately illustrates that
we may be seeing significant variability in our sample.

A more formal way to assess variability is to perform a
$\chi^2$ test on $N$ measurements of a quantity $x$ with error
$\sigma$ relative to the null-hypothesis of non-variability
using the equation
\begin{equation}
\chi^2 = \sum_i (\frac{x_i - \bar{x}}{\sigma_i})^2
\end{equation}
and $N-1$ degrees of freedom.  This measures the probability 
that the sample does not represent a constant variable with
normally distributed deviations from the mean, which we call the
variability probability.  We performed this test not only using
the measurement uncertainties reported in Tables 6 through 9,
but also for the possibility that our uncertainties are
underestimated by 15\%, as discussed in \S\ 3.7.  These results
are given in Table 11 and will be discussed on a star-by-star
basis.

We can use similar ideas to determine the statistical
significance of the difference between two measurements of a
value $x$ with error $\sigma$, using the Z-statistic:
\begin{equation}
Z = \frac{x_1 - x_2}{\sqrt{\sigma_1^2 + \sigma_2^2}}.
\end{equation}
We use this test mostly for long-term variability, but also
occasionally in the short-term dataset.

In Figures 4--6, we show the time series for equivalent width,
central wavelength, and linewidth for each of the four stars.
For convenience, we have expressed the central wavelengths as a
velocity relative to the average D$_3$ wavelength in the sample,
5875.72 \AA , consistent with the expected central wavelength of
the multiplet.  This will make it easier to assess whether any
apparent wavelength shifts are a large fraction of the
rotational velocity of the stars.

Before discussing each star, we note that since there are
several telluric lines within the D$_3$ profiles, if our removal
process does not work well we might see a dependence of the
equivalent width on the strength of telluric absorption during
that particular exposure.  Figure 7 shows this data for each
star.  The ``telluric line depth'' corresponds to the strongest
line in the vicinity of the sodium D lines, which is much
stronger than the lines that interfere with D$_3$.  Clearly, we
see no statistically significant trend of D$_3$ line strength
with telluric line strength.

In the following sections, we discuss the four stars from least
to most evidence of short-term variability in the D$_3$ line
strength.

\subsection{$\rho$ Gem}
This star falls well within the usual temperature and spectral
class range for early F-type stars and lies at the bottom of the
{\it Hipparcos} main sequence in Figure 2.  It has an M-type
companion which is 8 magnitudes fainter visually \citep{woo70},
so it does not contribute light to our spectra.  \citet{nor04}
report large radial velocity variability with a standard
deviation of 15.0 km s$^{-1}$ based on 3 measurements covering
3.1 years.  Our data cover a 5.0-year period and show at most
slight evidence for low-level long-term variability relative to
the $\sim$0.5 km s$^{-1}$ uncertainties, and our velocities are
similar to others referenced by SIMBAD.

We observed the star twice per night during each of the 8 nights
of the sp97 run.  In this dataset, the equivalent width, central
wavelength, and linewidth show weighted standard deviations that
are about 10--25\% greater than the mean uncertainties of the
original values.  Thus, as Table 6 indicates, a claim of
variability requires that our uncertainties are accurately
specified.

The time series in Figure 4 shows that there is one observation
with a small equivalent width (number 10 in Table 6), and that
both observations from the 6th night (numbers 13 and 14 in Table
6) are relatively high.  These three points provide almost
exactly half of the total $\chi^2$, and even with our reported
measurement uncertainties, the $\chi^2$ probability is not at a
level where we would claim variability.

The variability probabilities are greater for both central
wavelength and FWHM.  The central wavelength does not show a
particular pattern, but rather a scatter that appears larger
than the uncertainties while remaining a small fraction of $v$
sin $i$.  The FWHM mostly follows the pattern seen for
equivalent width, but with larger differences between the two
nightly measurements for several nights.

Overall, we do not find strong evidence for variability in the
strength of the D$_3$ line in $\rho$ Gem.  The evidence is
slightly stronger for the central wavelength and linewidth, but
requires that our measurement uncertainties are not
underestimated.

\subsection{$\theta$ Boo}
This is clearly a late F-type star that falls under the category
of ``solar-type,'' i.e., it is on the cool side of the point
near spectral type F5 where stars begin to show activity levels
related to the strength of the magnetic dynamo via rotational
speed and age \citep[e.g.][]{sim89,wol97}.  As Figure 2 shows,
this star appears to be about twice as luminous as a zero-age
main sequence star of the same color, thus it is either near the
end of its main sequence lifetime or is a binary system with
nearly equal components.  We see some evidence for radial
velocity variations in our data, and \citet{nor04} indicate both
radial velocity variability and an advanced age for this star
(3.1 Gyr).  However, we did not detect spectral lines from a
secondary star, and presume that any companion is faint.
Another possibility is that the two stars show minimal velocity
separation due to their orbit, e.g., a small inclination.  In
that case, the two stars would have to have very similar
spectral types, abundances, and rotational velocities.  Speckle
interferometry by \citet{mca92} constrains any companion to a
separation of less than 0.03 and $\Delta m$ greater than 1.5
mag.  For our purposes, this star was chosen due to small $v$
sin $i$, strong D$_3$ line, and its brightness, which gave us
the most precise data of the four stars with integration times
of 20 minutes.

We obtained repeated observations of this star in two different
observing runs.  In the sp97 run, we obtained 7 spectra over the
course of 5 consecutive observing nights, while in the sp98 run,
we obtained 12 spectra across 6 of the 7 nights.  Figures 4--6
show both of these time series.

We only see limited evidence for variability in equivalent width
or FWHM in both runs.  The $\chi^2$ value and resulting
probability for equivalent width variability in the sp98 run is
fairly large, but most of the total $\chi^2$ is due to a single
point, number 17 in Table 7.  Still, we see nothing unusual
about the spectrum or the fit that would invalidate this point.
If we compare this value with that from the 18th exposure taken
4.6 hours later, the Z-statistic gives a 3.3$\sigma$ difference,
corresponding to only a 0.1\% probability that these two values
result from a purely statistical variation.

The $\chi^2$ values indicate strong support for variability in
the central wavelengths, particularly in the sp97 run.  Figure 5
clearly shows the putative variation, which covers a range of
about 5 km s$^{-1}$ as compared with $v$ sin $i$ = 31.8 km
s$^{-1}$.  As noted in Table 5, we calculate a maximum possible
rotational period ($P$/sin $i$) of 2.8 days for this star.
Thus, the combination of a statistically constant D$_3$
equivalent width with an apparent oscillation in the central
wavelength over a few days would be consistent with rotational
modulation of one or more strong active longitudes analogous to
the Sun.  The low amplitude of the velocity variation in the
line would imply that the overall activity was broadly
distributed across the disk.  While this is a plausible
hypothesis, the data from the sp98 run do not as clearly show
this pattern, although the central wavelength does appear to be
variable.

\subsection{$\mu$ Vir}
As noted in \S\ 1, there is some evidence in the literature for
chromospheric variability.  As seen in Figure 2, this star is
located at the top of the {\it Hipparcos} main sequence,
consistent with being slightly evolved, although our derived
surface gravity is similar to main sequence stars.  It is also
the brightest star of the short-term variability sample and thus
we were able to obtain 26 observations during the 8 nights of
the sp98 run with typical integration times of 15 minutes.  As
with $\rho$ Gem, \citet{nor04} report significant radial
velocity variability with a standard deviation of 12.4 km
s$^{-1}$ in 3 measurements covering 1.4 years.  \citet{abt76}
reported possible variability of about 10 km s$^{-1}$ in 21
measurements spread over 4.8 years with a possible period of 0.9
years.  However, our measurements over 4.1 years (with intervals
between observations of 1.2, 1.3, and 1.7 years) only support
the possibility of a years-scale variation of up to 2 km
s$^{-1}$, a much smaller range than either previous dataset.

The $\chi^2$ analysis provides strong evidence for equivalent
width variability in our sample.  This variation is most easily
seen in Figure 4 by comparing the first and fourth night's worth
of data.  Interestingly, the appearance in the time series is
that of a gradual rise in activity over the first four nights,
followed by a decrease.  However, the value of the maximum
possible rotational period ($P$/sin $i$) in Table 5 is just 2.2
days.  Thus, this trend represents a more gradual variation than
would be explained by rotation.  In addition to night-to-night
variability, the time series also indicates variability within a
single night, particularly on the final night.  We do not find
significant periodicities at either time scale with a
periodogram analysis.

The variability probabilities for central wavelength are
extremely close to 1.  This variability is evident in the time
series in Figure 5 and appears strongest on short time scales
within a night.  The total range of the line centers is 15 km
s$^{-1}$ which is still a relatively small fraction of the D$_3$
line width or $v$ sin $i$.  A key point is that the variability
occurs on a much shorter timescale than rotation.

In a statistical sense, the variability in the wavelength is
more pronounced than that for the equivalent width.  One
possibility is that if the activity is spread across the disk in
some way, a rapid redistribution in point-to-point activity
levels could affect the disk-averaged central wavelength without
affecting the equivalent width as much.  One test of this
hypothesis would be to look for subtle line asymmetries, but at
the S/N of our data we can not confirm this possibility.

\subsection{18 Boo}
As already noted, the literature values for the D$_3$ equivalent
width significantly disagree.  The star lies near the bottom of
the {\it Hipparcos} main sequence in Figure 2, and it is a
possible member of the Ursa Major Moving Group
\citep{sod93,kin03}, which would also imply a relatively young
age of 300--500 Myr.  Our radial velocities in Table 9 show no
evidence for variability nor did we locate other evidence to
support classification as a spectroscopic binary although it is
a wide visual binary.

This star shows very strong evidence for line strength
variability, the strongest in the present sample.  Figure 4
clearly shows the variability, and the probabilities in Table 11
are very large.  The variability on the 4th and 7th nights is
highly pronounced and covers a large fraction of the range seen
in the full sample.  In contrast, the line strength is tightly
clustered on the 5th night.  It is important to note that the
observed variability in 18 Boo on the 4th and 7th nights is not
matched by $\mu$ Vir on those nights.  For all spectra of the
two stars, if one star was observed the other star was observed
immediately before and/or after.  This gives further support to
the hypothesis that our observed variability is not a data
analysis artifact.  As with $\mu$ Vir, we do not find strong
evidence for periodicities in the time series.

To further illustrate the line variability, we show a stack plot
of all 25 spectra from the sp98 run in Figure 8, along with the
Gaussian fits.  The variability can be most clearly seen near
the top, as well as when looking closely at the variable spacing
between adjacent spectra at the line core, despite the uniform
spacing of the continua.

The variability probabilities for the linewidths are similar to
those for line strength.  A comparison of Figures 4 and 6
indicate that the putative variability in both quantities is
related and this relationship between equivalent width and FWHM
is most pronounced for 18 Boo.  The variability probabilities
are smaller for the central wavelength.  The total velocity
range of about 12 km s$^{-1}$ is covered by the spectra on the
first night, with a smaller range for the rest of the data.  As
with $\mu$ Vir, the implication is that the variability is
operating on non-rotational timescales.

Based on our measurements, we believe that the large difference
in the two previous literature values of equivalent width
represents the same variability that we have observed.  Indeed,
those measurements provide a good match to the high and low
values of equivalent width that we report.

As one final additional test of our D$_3$ results, we fitted
several photospheric lines to see if those lines appeared
variable.  Since 18 Boo shows the strongest evidence for D$_3$
variability, we give those results in Table 12, which are also
representative of $\mu$ Vir.  We see that for two lines that are
somewhat stronger than D$_3$ and one line that is comparable to
the weakest D$_3$ measurement, the standard deviation of the
line measurements is nearly identical to the measurement
uncertainties.  A $\chi^2$ analysis gives variability
probabilities generally in the range 0.4--0.8, supporting the
idea that our measurement techniques are not creating the
appearance of variability in the D$_3$ lines where none is
present.  Out of the nine probabilities, only one is greater
than 0.9, but that is for the central wavelength of the weakest
line, and only a few of our D$_3$ lines in 18 Boo are that weak.
Furthermore, a Gaussian is not as good of a match to the
photospheric lines as for the D$_3$ line and that might
contribute to larger fluctuations in the fits.

\section{Long-term variability}

\subsection{Preliminary comments}
We obtained additional measurements of the four stars in the
short-term sample to look for long-term variability and those
measurements have already been presented in Tables 6--9.  Table
13 provides measurements of an additional eleven stars for which
we obtained measurements in at least two different observing
runs and we give the same data as for the short-term sample.  In
the rest of this section we make note of literature values of
the D$_3$ equivalent width where available.  While one can not
exactly compare these results due to differences in the telluric
and photospheric line removal procedures, those values should at
least be similar to ours if there is no variability.  As with
the short-term study, we give heliocentric radial velocities.
Again, we want to make sure that the D$_3$ lines are not being
``polluted'' by a cooler star or that binarity is affecting the
activity levels.  In no case did we see any spectral lines from
a secondary component.

\subsection{Long-term variability in the short-term sample}
As seen in Tables 6--9, for all four stars in the short-term
sample we obtained spectra in at least three additional
observing runs.  These values are broadly consistent with the
short-term variability (or lack thereof) seen in the large
samples within a single observing run.

The evidence for equivalent width variability for both $\rho$
Gem and $\theta$ Boo becomes more significant when including the
additional data.  For $\rho$ Gem, three out of the four
additional spectra (numbers 1, 2, and 20 in Table 6) show
equivalent widths considerably smaller than the bulk of the sp97
sample.  In fact, these values match the single low point
(number 10) seen in that sample.  If we include all 20
measurements of $\rho$ Gem, the probability of equivalent width
variability rises to 0.994 with our reported uncertainties and
0.926 if our uncertainties are underestimated by 15\%.

Our two sets of observations for $\theta$ Boo indicate long-term
variability.  If we compare the values in Table 10, 41.1
$\pm$ 1.6 m\AA\ for sp97 and 33.3 $\pm$ 3.2 m\AA\ for sp98, we
derive a 2.2$\sigma$ difference using the Z-statistic, and that
does not take into account a $\sqrt{N}$ reduction in the
uncertainty if we used the sample standard deviation.  Notably,
all seven equivalent widths from sp97 are larger than all twelve
values from sp98.

Previous investigators have reported equivalent widths for all
four stars.  We have already mentioned that the previously
discrepant 18 Boo measurements are consistent with our observed
variability.  Previous values for $\rho$ Gem match the high end
of our measurements; 38 m\AA\ from \citet{wol86} and 7
measurements of 30.0--48.2 m\AA\ from \citet{ter05}.
\citet{wol86} found an average equivalent width for $\mu$ Vir of
19 m\AA\ from 12 measurements, somewhat smaller than our average
of 28.5 m\AA .  Finally, \citet{wol86} found an equivalent width
of 35 m\AA\ for $\theta$ Boo and \citet{ter05} found 27.7--36.2
m\AA\ in 5 measurements; these values are generally consistent
with ours.

\subsection{Additional stars in the long-term sample}

\subsubsection{$\beta$ Cas}
This star lies well above the main sequence in the Hertzsprung
Gap, and has by far the smallest surface gravity in our sample.
The star has one of the largest $v$ sin $i$ values in the
sample, and thus the radial velocities are particularly
uncertain and not indicative of variability.  We have already
mentioned the possible UV emission line variability in \S\ 1.

As found by \citet{rac97}, giant stars in this temperature range
show similar D$_3$ equivalent widths as dwarfs, and our D$_3$
equivalent widths for $\beta$ Cas are typical for early F-type
stars.  The first two equivalent widths are statistically
identical, but the third value differs from the second value at
the 2.0$\sigma$ level.

Interestingly, while the central wavelengths of the D$_3$ line
are not statistically variable, they consistently deviate to the
blue.  An average of the three measurements yields 5875.41 $\pm$
0.15 \AA , or a blueshift of 16 $\pm$ 7 km s$^{-1}$ relative to
the 5875.72 \AA\ mean found for the short-term sample.
\citet{ayr98} found similar blueshifts for the O IV] 1401, C IV
1548, and C IV 1550 lines and similar amounts of excess
broadening beyond the photospheric $v$ sin $i$.  These lines
form at temperatures near 10$^{5}$ K, while the D$_3$ line is
thought to form at temperatures near 10$^{4}$ K.  However, other
UV emission lines that form at temperatures comparable to D$_3$
did not show these blueshifts.

\subsubsection{9 Aur}
This star is one of the best-studied $\gamma$ Doradus non-radial
pulsating variables \citep{kri95}.  There is a 12th magnitude
M-type component 5 arcsec from the main star which is too faint
to affect our spectra.  Our first D$_3$ measurement is rather
uncertain, but no variability is detected.

\subsubsection{$\eta$ Lep}
This star lies very near 9 Aur on the {\it Hipparcos}
color-magnitude diagram in Figure 2 with similar temperature and
gravity, but appears not to show significant photometric or
radial velocity variability.  In particular, \citet{nor04}
report statistically constant radial velocity in 13 measurements
covering 16.3 years and our three measurements spanning 1.3
years agree with this conclusion.

The equivalent widths and line widths for the three observations
are statistically identical.  However, the central wavelength
for the first and second observations disagree at the
2.9$\sigma$ level based on the Z-statistic.  The wavelength
difference corresponds to a velocity difference of 3.9 $\pm$ 1.2
km s$^{-1}$, a small fraction of $v$ sin $i$ = 17.4 km s$^{-1}$.
This implies a situation similar to that for $\mu$ Vir where
there may be some sort of redistribution of activity that causes
little if any equivalent width variability, but slightly
distorts the shape of the line leading to a change in the
measured line center.

\subsubsection{$\alpha$ CMi (Procyon)}
Procyon was observed not only because it is the brightest main
sequence F-type star, but also because at spectral type F5 and
$B-V$ = 0.42 it is at the high temperature edge of solar-type
stars which show activity-rotation-age correlations
\citep[e.g.][]{sim89,wol97}.  This star exhibits low-amplitude
variability in the Ca II H \& K lines \citep{bal95}.
\citet{dan85} found a D$_3$ equivalent width of 5 m\AA\ and our
three values are similar and highly consistent, as are the line
widths.

In contrast, the central wavelength of the second observation is
considerably bluer than the other two measurements.  The
Z-statistic indicates that the second and third measurements
differ at the 2.8$\sigma$ level.  Given that this appears to be a
solar-type star, the situation may be similar to what we found
for $\theta$ Boo, a variation in wavelength due to rotational
modulation of active regions.  It is important to note that the
first two observations were made just 5 days apart and $P$/sin
$i$ for Procyon is 18.5 days, thus our observations likely cover
something close to one-quarter of a full rotation.  The
wavelength difference between these two observations corresponds
to 12 $\pm$ 7 km s$^{-1}$.  Recall that we found $v$ sin $i$ =
5.3 km s$^{-1}$, which would also be the maximum observable
velocity range for one-quarter of a period.  \citet{all02}
derive $v$ sin $i$ = 3.2 km s$^{-1}$ from a more
sophisticated 3-dimensional model atmosphere.  Thus, our results
are statistically consistent with rotational modulation, with
the caveat that our first measurement is rather uncertain.

\subsubsection{$\chi$ Leo}
This star is an interesting case due to the very weak D$_3$ line
for an otherwise normal, but slightly evolved early F-type star.
We see evidence for long-term radial velocity variability, but
this mostly appears in the sp95 observation for which there may
be a slight zero-point offset relative to other runs due to the
more limited wavelength coverage.  We did not find significant
evidence for radial velocity variability in the literature.

For the four spectra in which we could measure the D$_3$ line at
the 2$\sigma$ level or greater, the measurements are statistically
identical.  We were unable to obtain a statistically significant
D$_3$ detection in the sp98 data, but given the weakness of the
line in the other spectra, this non-detection is consistent with
the other measurements.

Although somewhat uncertain, the linewidths are relatively
narrow.  The weighted mean and weighted standard deviation for
the four D$_3$ detections give FWHM = 35 $\pm$ 13 km s$^{-1}$,
as compared with $v$ sin $i$ = 27.5 km s$^{-1}$.  Although
$\sigma$ Boo (see below) has a very narrow D$_3$ line, it also
has very small $v$ sin $i$, and $\chi$ Leo is the only star in
our present sample with an average D$_3$ linewidth only slightly
broader than photospheric lines.  For comparison, $\alpha$ CMi
also shows a consistently weak D$_3$ line, but with FWHM =
56--57 km s$^{-1}$, indicating the effects of greater thermal
and turbulent broadening for D$_3$ as compared with photospheric
lines.

In principle, a small equivalent width for a D$_3$ line in an
F-type star suggests a low active-region filling factor
\citep{and95}, although the hottest star considered in that
study was 6500 K.  In turn, a small active-region filling factor
would generally produce a narrower line than the case of a
larger filling factor, particularly if the activity were
constrained to a limited longitude range and/or high latitudes.
\citet{rac00} found one other case of an unusually narrow D$_3$
line in a Pleiades star (H II 1266), but in that case the D$_3$
line was clearly narrower than $v$ sin $i$ even when not
compensating for the larger thermal and turbulent broadening for
D$_3$.  Activity concentrated at high latitudes was invoked to
explain the contradiction in linewidths.  In that case, $v$ sin
$i$ = 76.6 km s$^{-1}$, so it was much easier to rule out
spectrum-to-spectrum wavelength shifts in the D$_3$ line to
within very small fraction of rotational broadening and thus
mostly rule out rotational modulation of an active region at a
single stellar longitude.  Since $\chi$ Leo has one-third as
much rotational broadening as the Pleiades star, our wavelength
uncertainties amount to 10--20\% of $v$ sin $i$.  Thus, even
though our D$_3$ central wavelengths show no variability within
the precision of our measurements, we can not as easily rule out
rotational modulation.

\subsubsection{$\alpha$ Crv}
This star lies at the bottom of the {\it Hipparcos} main
sequence, although the spectral classification indicates a
slightly evolved star.  The heliocentric radial velocity
difference between our two observations would be statistically
significant, but as already mentioned, we have concerns about
the zero-point of the sp95 observations.  \citet{nor04} find a
strong likelihood of radial velocity variability in 20
measurements over 14.9 years.  We see no significant difference
between the two D$_3$ measurements.

\subsubsection{$\sigma$ Boo}
This star lies at the bottom of the {\it Hipparcos} main
sequence.  \citet{nid02} found radial velocity scatter less of
than 0.1 km s$^{-1}$ based on high-precision observations used
in a search for exoplanets.  Note that our velocities are about
1 km s$^{-1}$ greater than theirs, which gives some indication
of the systematic errors that may exist in our radial velocity
measurements.

The two D$_3$ equivalent widths differ by 2.4$\sigma$ via the
Z-statistic.  The literature values are consistent with the
possibility of variability as \citet{wol86} found an equivalent
width of 8 m\AA , while \citet{ter05} found 19.0 m\AA .  Our
central wavelength and linewidth measurements are not
statistically different.

\subsubsection{$\mu^1$ Boo}
This star has the largest $v$ sin $i$ in our sample.  It has a
wide companion nearly 2 arcmin away which is itself a
long-period binary ($\mu^2$ Boo).  Photographic spectra indicate
at best low-amplitude radial velocity variability
\citep{nie70,abt74}.  Our data is also of relatively low
precision due to the large $v$ sin $i$ and low S/N for the
observation that appears to differ from the other two.

The two observations in which we detect D$_3$ show reasonable
statistical agreement.  We do not conclusively detect D$_3$ in
the low quality observation.  However, depending on the width of
a putative D$_3$ line, the 2$\sigma$ upper limit could be
comparable the other observations.

\subsubsection{$\sigma$ Ser}
Our D$_3$ measurements are statistically identical and given the
relatively large $v$ sin $i$ and the possible zero-point issue
with the sp95 data, the heliocentric radial velocities are
likely non-variable.  \citet{wol86} found an equivalent
width of 23 m\AA\ and \citet{ter05} found 29.2 m\AA , broadly
consistent with our values.

\subsubsection{HR 6237}
This star is a single-line spectroscopic binary with a period of
3.8 years \citep{abt74}.  Our two radial velocity
measurements are broadly comparable to those results.  The
wavelength of the D$_3$ line differs by 2.2$\sigma$ between the
two observations, but the line strength and width are nearly
identical.

\subsubsection{$\xi$ Oph}
This star is a visual binary with $\Delta m$ = 4.0 and a
separation of 14.4 arcsec.  Our radial velocities suggest
variability, as do the three observations by \citet{nor04}
covering 263 days.  However, the D$_3$ measurements are nearly
identical in all three parameters.

\section{Discussion}
Our results for 18 Boo and to a lesser extent $\mu$ Vir appear
to establish the reality of short-term chromospheric variability
in some fraction of early F-type stars.  Both stars appear to be
single with no reported optical variability in {\it Hipparcos}
observations or other literature sources.  Stellar parameters
are quite similar for the two stars, although $\mu$ Vir appears
to be somewhat evolved towards the terminal-age main sequence
while 18 Boo appears to be very near the zero-age main sequence,
consistent with its possible membership in the Ursa Major Moving
Group.

The physical nature of this variability is not clear, but we can
explore various situations involving rotation.  While we do not
know the sin $i$ factors for individual stars, the maximum
possible value of $v$ should be similar to or less than the
largest $v$ sin $i$ seen in young clusters.  For the Pleiades
and Alpha Persei clusters, \citet{kra67} found early F-type
stars with $v$ sin $i$ $\sim$ 200 km s$^{-1}$.  Even in that
extreme case, the rotational period would still be $\sim$8 hours
and pure rotational modulation of isolated active regions would
not seem to be responsible for the variability we see.
Furthermore, even in the early F-type stars, rotation slows down
with age and with decreasing effective temperature.  In fact,
for the temperature range of our field stars, $\langle v$ sin
$i\rangle$ is about 30--50 km s$^{-1}$ \citep{wol97}.  Thus, the
three early F-type stars in the short-term sample have nearly
average values of $v$ sin $i$ and are not likely to be viewed
nearly pole-on.  Thus the rotational periods will be much closer
to the values in Table 5 than they are to the hypothetical
minimum value of $\sim$8 hours.  We have already noted that we
do not find any clear periodicities in the equivalent widths,
albeit the temporal coverage is limited.

Furthermore, the D$_3$ line in this temperature range is purely
chromospheric, e.g., in the Sun the line is only seen in
conjuction with plages \citep{lan81}.  Thus, the line only
samples the portions of the rotational broadening function at
speeds corresponding to the apparent velocities at those stellar
longitudes.  If we were seeing rotational modulation of a small
number of isolated active regions, we should see velocity shifts
(or asymmetries) in the D$_3$ line that are a large fraction of
$v$ sin $i$, unless the active regions are all at high
latitudes.  However, Figure 5 clearly shows that such large
shifts are not observed.  The total velocity range of the D$_3$
line for both 18 Boo and $\mu$ Vir is less than 20\% of the
quantity 2 $\times$ $v$ sin $i$.  Moreover, the rotational
modulation produced by high-latitude features would be less
likely to explain the full range of equivalent width variation
that is seen for these two stars.

Another aspect of significant axial inclination is that we would
mostly be observing polar regions.  In the Sun, activity is
concentrated near the equator, thus in other stars with sun-like
activity, over time we should see activity covering the full
range of equatorial rotation.  However, the other consequence of
observing polar regions is the possibility of significant
stellar oblateness.  Interferometric studies of rapidly rotating
($\sim$200 km s$^{-1}$) A-type stars indicates that such stars
can be highly oblate, which causes the surface temperature to be
considerably hotter in the polar regions \citep[e.g.,][]{van01}.
Thus, a star may be cool enough at the equator for convective
activity while being hot enough near the poles to shut off
convection.  This effect could influence the observed strength
and distribution of active regions on the star and thus the
appearence of chromospheric spectral features
\citep[e.g.,][]{fre95}.  However, we have no evidence that the
stars in our sample are rotating at such high speeds.  Plus,
this effect would merely mimic the observed situation on the Sun
where activity tends to lie near the equator.

Our results suggest that the observed variability must either
result from some sort of global phenomenon or significant
variations of a large number of small active regions.  Either of
these possibilities could produce significant short-term
activity variations without large velocity variations or obvious
periodicity, and thus no observable relationship with rotation.

As a final note on short-term variability, it is worth
considering in more detail the actual distribution of equivalent
width values in our sample.  We have already shown with the
$\chi^2$ analysis that the equivalent widths for both 18 Boo and
$\mu$ Vir significantly differ from a Gaussian distribution of
width comparable to the measurement uncertainties.  In Figure 9,
we give histograms of the equivalent widths for both stars.  If
changes in activity were mostly the result of some level of
flaring on top of a basal level of activity, we might expect
many points for a star to be clustered in the low end of the
distribution with a noticeable ``tail'' of larger equivalent
widths.  If the activity were somehow related to a ``high/low
state'' phenomenon we would expect a bimodal distribution.  (A
similar result would occur if we were seeing a purely sinusoidal
variation whereby the star would spend more time near the peak
and trough of the variation.)  However, we see nothing
conclusive in the distributions.  For both stars the
distributions look like something in between a broad Gaussian
and a uniform distribution, with just a couple of high points
for 18 Boo.  Admittedly, with our limited number of data points,
the $\sqrt{N}$ errors are large in the histograms, but the point
is that we do not see a clear pattern that would indicate a
well-defined mode of variability.

As for the long-term variations, we can only claim 2$\sigma$
variability in the D$_3$ equivalent width for two of the eleven
stars in the purely long-term sample, $\beta$ Cas and $\sigma$
Boo.  Given the apparent short-term variability we have found in
$\mu$ Vir and 18 Boo, it may be difficult to claim long-term
variability based on a small number of measurements because a
significant difference between two measurements widely separated
in time might simply be due to the short-term variability.  

However, we can use the short-term results to set an expectation
on how often we might find such variability in the long-term
sample, even if there is no true long-term variability.  We can
take all combinations of the short-term observations for a star
and see how often we get a Z-statistic greater than 2.0 or 3.0.
For $\rho$ Gem, there are 120 possible pairs of observations
that can be chosen, 14 (11.7\%) of which have Z $>$ 2.0,
including just 2 (1.7\%) with Z $>$ 3.0.  In contrast, for 18
Boo there are 300 possible pairs of observations, 88 (29.3\%) of
which have Z $>$ 2.0, including 29 (9.7\%) with Z $>$ 3.0.
Naturally, we can also note that if all stars are constant, we
expect only 4.6\% of a large sample to show Z $>$ 2.0 and 0.3\%
with Z $>$ 3.0.

Our finding of only two out of eleven stars with possible
long-term variability is thus most consistent with the minimal
or non-existent short-term variability of $\rho$ Gem.  However,
given the small sample size, this finding is also consistent
with the other two possibilities.  The key is that our results
do not suggest significant excess long-term variability in the
early F-type stars beyond that which can be explained by
short-term variability, unless the long-term variability happens
on very long time scales.  If long-term variability were
important on several year time scales, we might expect a large
fraction of our long-term sample to show variations due to a
combination of short-term and long-term variability and that
appears not to be the case.

However, we do not have enough information to support an
explanation for why we do not see variability in more stars.
Given the long-term data for $\rho$ Gem, the combination of bad
luck and slightly poorer data quality may have conspired to hide
more obvious short-term variability.  In fact, we certainly can
not rule out the possibility that all early F-type stars exhibit
variability at a level that would be detected by a large number
of observations at a precision comparable to our best data.  It
is worth noting that most stars in the long-term sample have
average line strengths below that of the short-term sample,
making it more difficult to detect activity variations at a
given fractional level.

As far as the activity levels themselves, it is clear from a
variety of studies using numerous activity indicators that the
early F-type stars exhibit a wide range in activity, even when
considering a narrow temperature range.  This is true for
chromospheric emission lines such as \ion{C}{2} $\lambda$1335
\citep{sim91} and Lyman $\alpha$ \citep{lan93}, and X-ray
emission \citep{sch85}.  Results for the D$_3$ line are similar
\citep{gar93,rac97,rac00}.  However, the results of these
studies can rule out some mechanisms that might explain the
range, such as luminosity class \citep{rac97}, age
\citep{rac00}, and rotational velocity \citep{sim91,rac97}.

An important motivation for the present study was to assess
whether short-term and/or long-term variability can explain the
large range in activity levels.  This does not appear to be the
case.  Among the most persuasive arguments to support this
conclusion are the general lack of significant variability in
the long-term sample, and the fact that $\chi$ Leo seems to have
consistently very low activity.  Stars such as $\sigma$ Ser and
$\eta$ Lep are near $\chi$ Leo on the color-magnitude diagram
and have very similar temperatures and gravities, but have
D$_3$ line strengths a factor of $\sim$5 greater.  Thus, if
variability is responsible for this difference, it must happen
on time scales longer than a few years.  Furthermore, the
general agreement we see between our measurements and those
reported by other authors many years before and/or after also
suggests that long-term variability is not a major factor.

\section{Conclusions}
We have performed the most detailed study to date on
chromospheric variability in the early F-type stars.  Through a
combination of intensive observations of four stars over
intervals of several days and occasional observations over a
period of a few years for these stars plus an additional eleven
stars, we have searched for variations in the strength,
wavelength, and width of the helium D$_3$ line.  Key aspects of
our study are the rigorous procedures used to eliminate
contaminating telluric and photospheric lines from the
chromospheric D$_3$ line and our careful assessment of
measurement uncertainties.

We find significant evidence for short-term (hours to days)
variability in two early F-type stars, amounting to about a
factor of two in equivalent width.  The central wavelength of
the line also shows evidence for variability, but this
variability covers a small fraction of the total range due to
rotational broadening.  Our data do not support a simple
explanation associated with pure rotational modulation of
discrete active regions or active longitudes for the early
F-type stars, but this explanation does appear to apply to the
short-term variations seen in the two coolest stars, $\alpha$
CMi and $\theta$ Boo.

In a statistical sense, the small number of stars in the
long-term sample showing possible variability is consistent with
the idea that some small fraction of early F-type stars show
short-term variability large enough to be detected at our
measurement precision.  However, the long-term sample does not
point to the likelihood of variability on the scale of years in
the early F-type stars beyond that which can be explained by
short-term variability.

Finally, the general lack of variability larger than a factor of
two implies that variability is not an explanation for the large
range in activity levels seen in the early F-type stars and this
range remains unexplained.

\acknowledgments
We thank the referee for very helpful comments on the
manuscript.  We would also like to thank Darrel Smith for his
careful reading of the manuscript and Andri Gretarsson for
discussions concerning statistical analysis.  This research has
made use of the SIMBAD database, operated at CDS, Strasbourg,
France.  Participation by DRF was supported by a NASA
traineeship grant provided by the Arizona/NASA Space Grant
Consortium.

\clearpage
\begin{figure}
\plotone{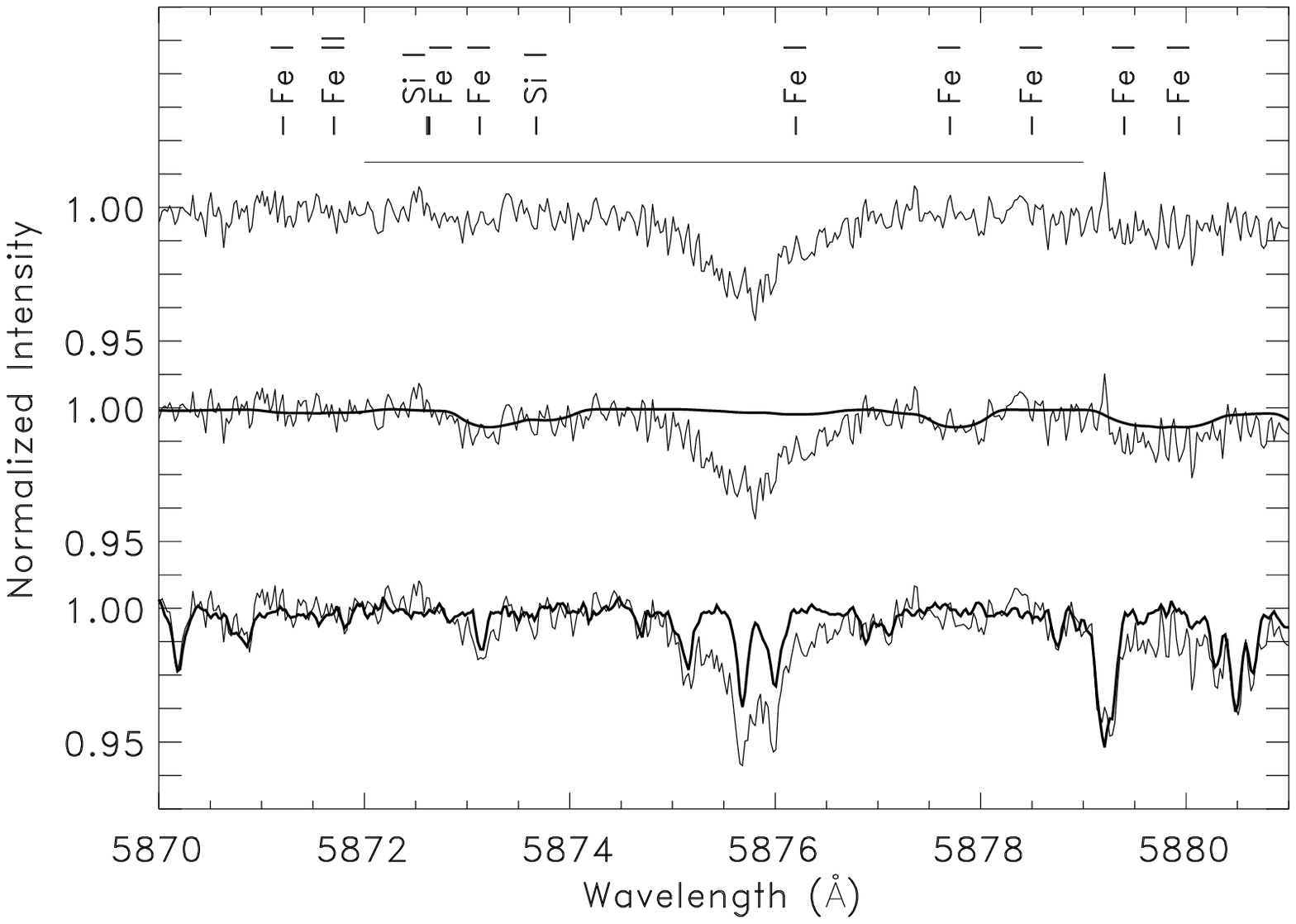}
\caption{Sample spectrum for the D$_3$ region showing the
sequence of corrections.  Bottom: normalized raw spectrum for
$\eta$ Lep with telluric model overplotted (telluric absorption
is relatively strong in this spectrum, while photospheric line
strengths are about average).  Middle: telluric-corrected
spectrum with photospheric model overplotted.  Top: final D$_3$
spectrum, with solid line depicting the fit region used to
derive line parameters; note that the actual D$_3$ fit is
performed on an unnormalized spectrum.}
\end{figure}

\clearpage
\begin{figure}
\plotone{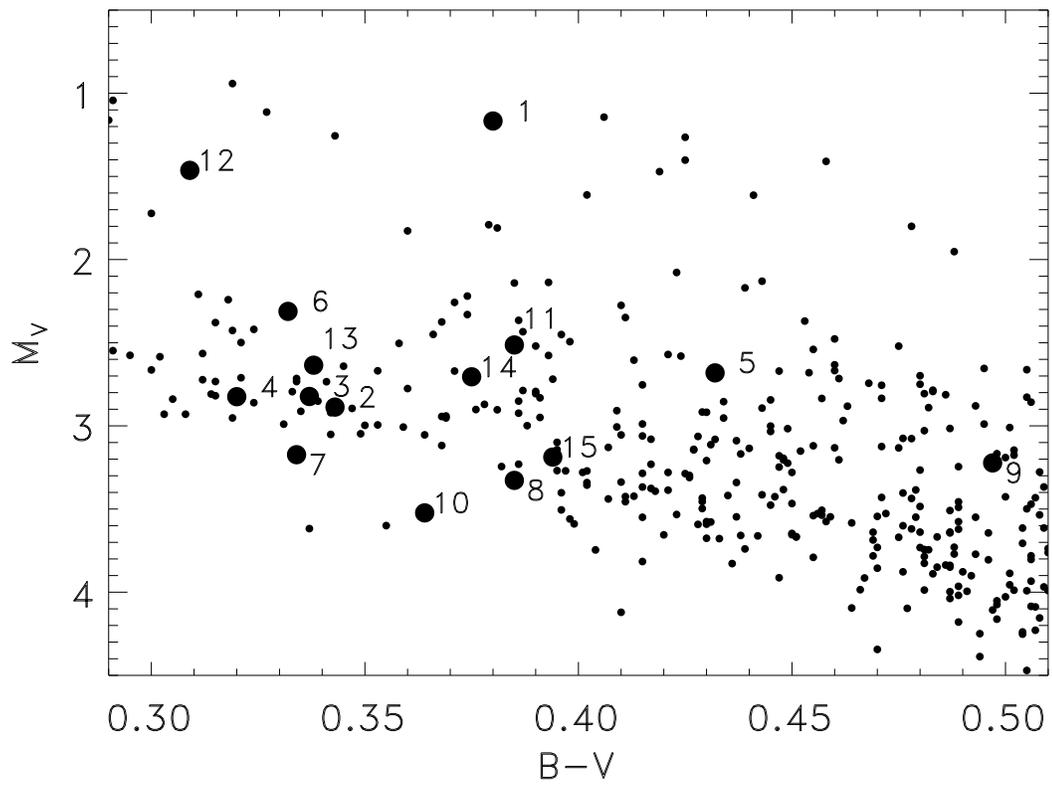}
\caption{Color-magnitude diagram for stars with {\it
Hipparcos} parallax greater than 25 mas.  Program stars are
numbered to correspond to Tables 4 and 5.}
\end{figure}

\begin{figure}
\plotone{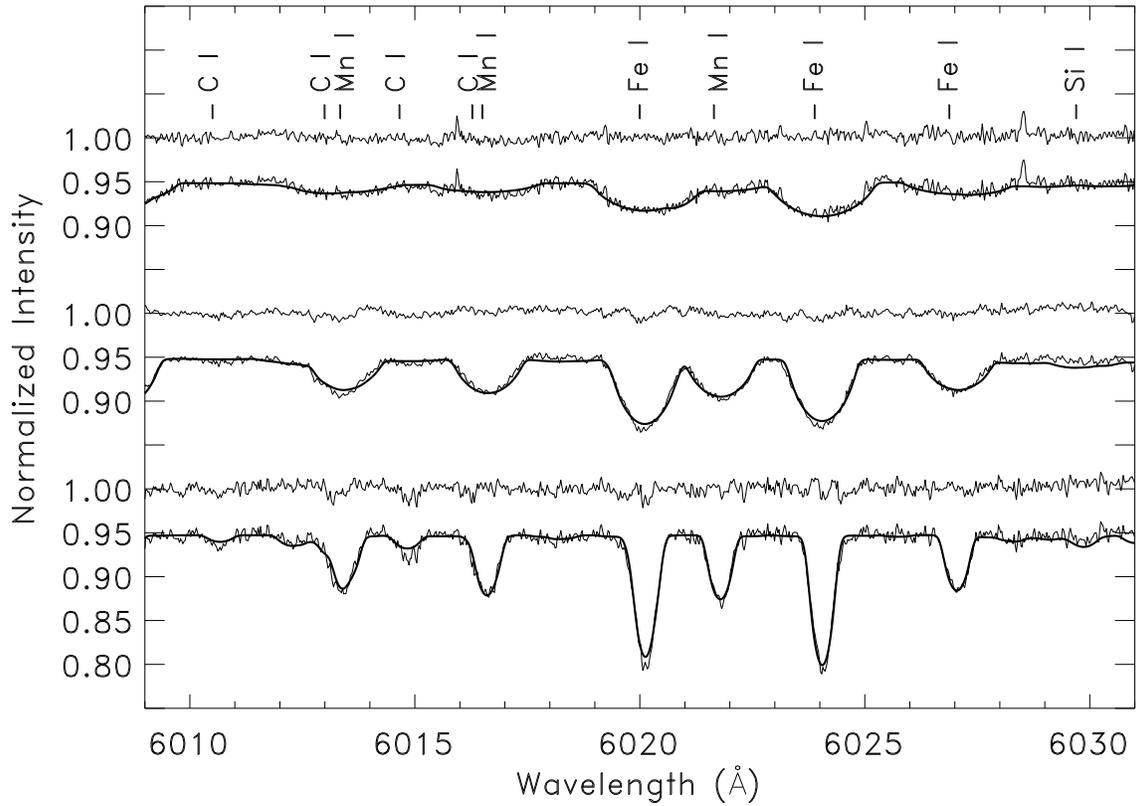}
\caption{Comparison of observed (thin lines) and
synthetic (thick lines) spectra for a range of rotational
broadening for a portion of the 6007--6200 \AA\ fit range.
Residuals in the sense observed divided by synthetic are given
above each spectrum.  From top to bottom, stars are $\rho$ Gem
($v$ sin $i$ = 58.3 km s$^{-1}$), 18 Boo (38.9 km s$^{-1}$), and
$\eta$ Lep (17.4 km s$^{-1}$).  Species identifications are
given for the strongest lines.}
\end{figure}

\clearpage
\begin{figure}
\plotone{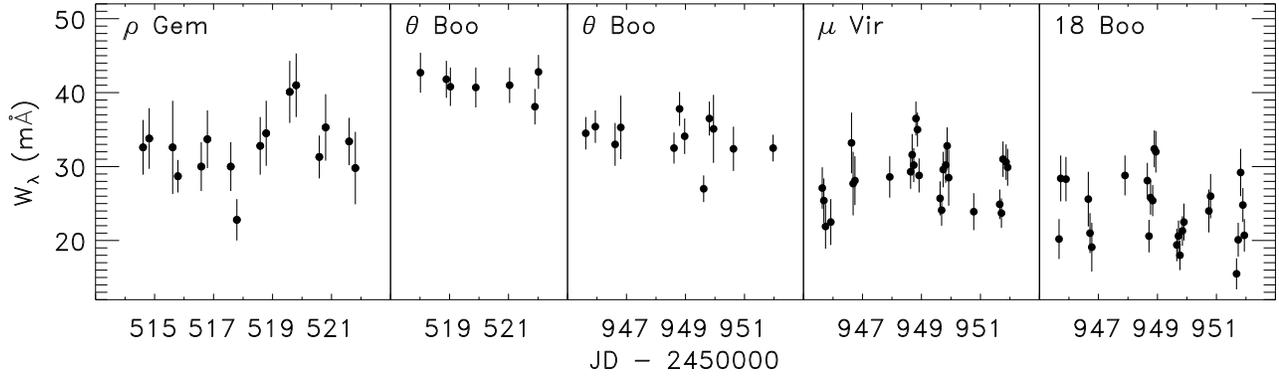}
\caption{Time series of D$_3$ equivalent widths.}
\end{figure}

\begin{figure}
\plotone{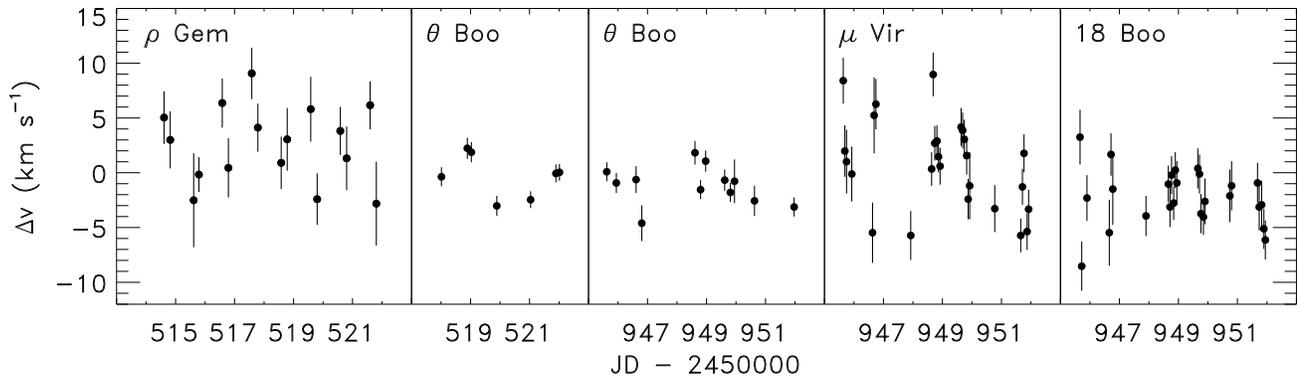}
\caption{Time series of D$_3$ line velocities, relative to the
sample mean.}
\end{figure}

\begin{figure}
\plotone{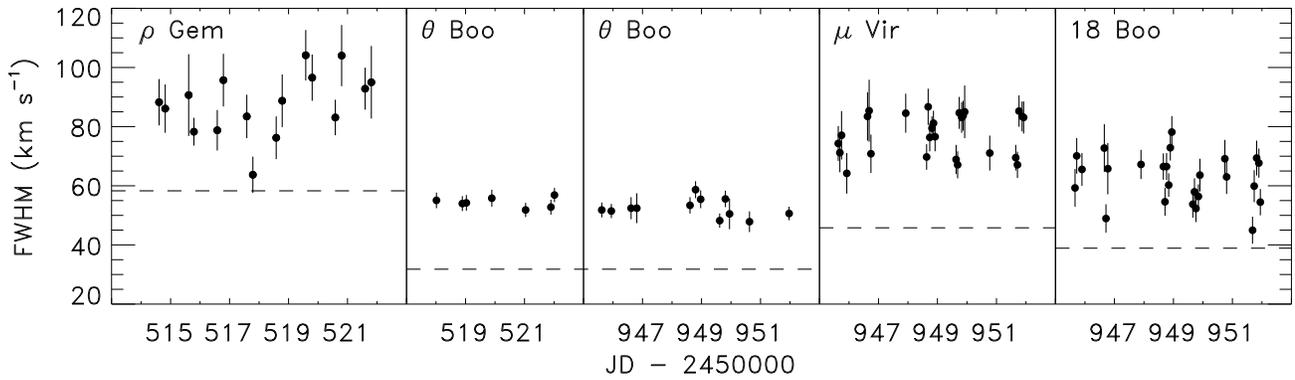}
\caption{Time series of D$_3$ linewidths.  For comparison,
stellar $v$ sin $i$ is indicated by the dashed lines.}
\end{figure}

\begin{figure}
\plotone{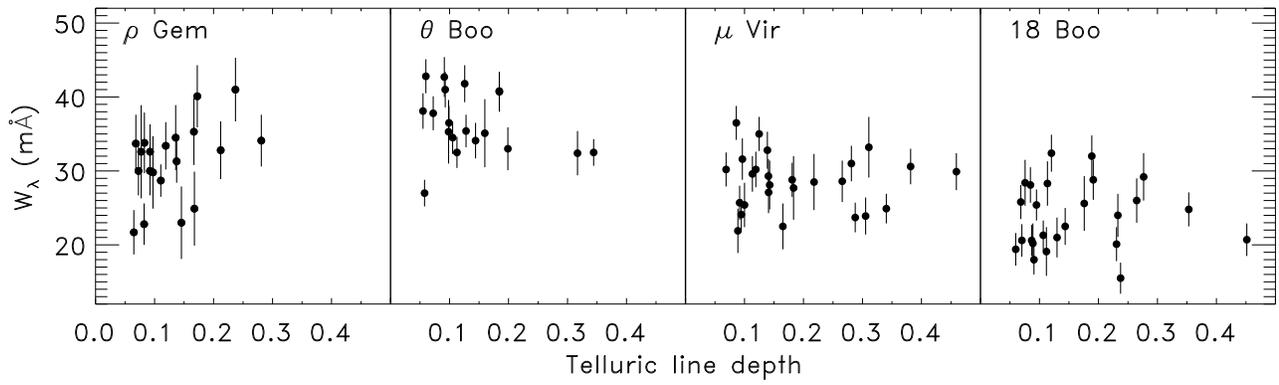}
\caption{D$_3$ equivalent width versus telluric line strength.
The latter is based on the strongest telluric line in our fit
range.}
\end{figure}

\clearpage
\begin{figure}
\epsscale{0.50}
\plotone{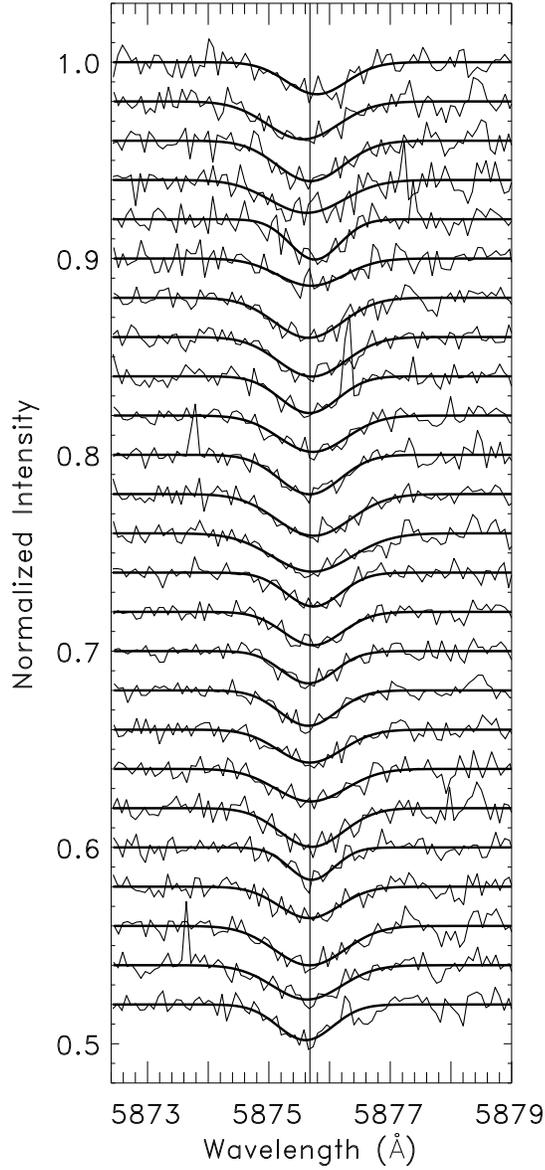}
\caption{D$_3$ spectra and fits for 18 Boo for the sp98 dataset.
All 25 spectra are shown in chronological order from top to
bottom.  The occasional upward spikes in the data are due to
radiation events in the CCD and carry negligible statistical
weights in the line fits.  Significant variations are visible,
especially near the top.}
\end{figure}

\clearpage
\begin{figure}
\epsscale{1.0}
\plotone{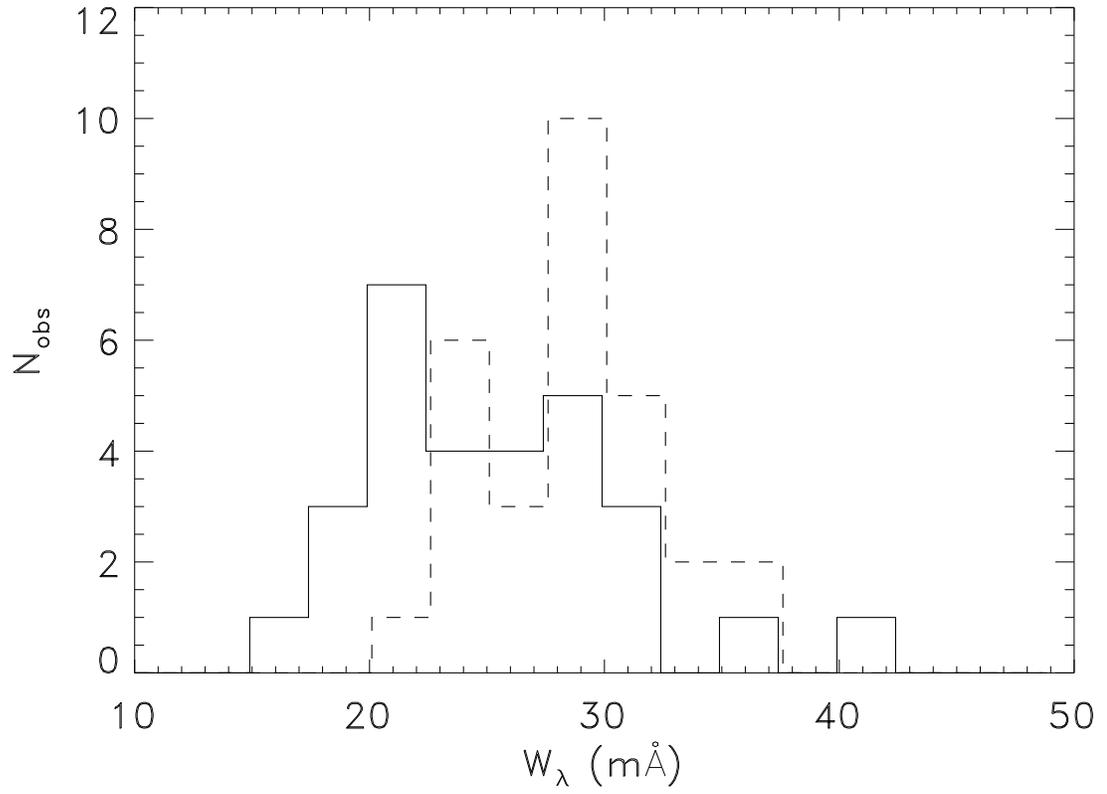}
\caption{Histograms of D$_3$ equivalent widths.  Solid line
corresponds to 18 Boo and dashed line corresponds to $\mu$ Vir.
The binsize of 2.5 m\AA\ was chosen as an approximate match to
the typical measurement uncertainties, and the two histograms
are slightly offset horizontally for clarity.}
\end{figure}

\clearpage
\begin{deluxetable}{ccc}
\tablecaption{Observing runs}
\tablewidth{0pt}
\tabletypesize{\footnotesize}
\tablenum{1}
\tablehead{
  \colhead{Run} & \colhead{Dates} & \colhead{Wavelengths} \\
& \colhead{(UT)} & \colhead{(\AA )}
}
\startdata
sp95 & 1995 Feb 2--8 & 5781--5980 \\
fa95 & 1995 Nov 28--Dec 6 & 5781--6844 \\
sp96 & 1996 Feb 20--29 & 5552--6928 \\
fa96 & 1996 Oct 17--22 & 5781--6243 \\
sp97 & 1997 Mar 7--14 & 5781--6243 \\
sp98 & 1998 May 12--18 & 5781--6243 \\
fa99 & 2000 Jan 19--27 & 5661--6840
\enddata
\end{deluxetable}

\begin{deluxetable}{cccc}
\tablecaption{Solar abundance fit}
\tablewidth{0pt}
\tabletypesize{\footnotesize}
\tablenum{2}
\tablehead{
  \colhead{Value} & \colhead{Present} &
  \colhead{Gray\tablenotemark{a}}
& \colhead{GS\tablenotemark{b}}
}
\startdata
log A(C)   &8.52  &8.66  &8.52  \\
log A(O)   &8.94  &8.91  &8.83  \\
log A(Si)  &7.61  &7.64  &7.55  \\
log A(S)   &7.21  &7.23  &7.33  \\
log A(Ca)  &6.37  &6.34  &6.36  \\
log A(Mn)  &5.56  &5.42  &5.39  \\
log A(Fe)  &7.54  &7.59  &7.50  \\
log A(Ni)  &6.31  &6.27  &6.25  \\
log A(Ba)  &2.19  &2.13  &2.13
\enddata
\tablenotetext{a}{Values from \citet{gra88}}
\tablenotetext{b}{Values from \citet{gre98}}
\end{deluxetable}

\begin{deluxetable}{cccc}
\tablecaption{Procyon abundance fits}
\tablewidth{0pt}
\tabletypesize{\footnotesize}
\tablenum{3}
\tablehead{
  \colhead{Value} & \colhead{6625\tablenotemark{a}} & \colhead{6750\tablenotemark{a}}
  & \colhead{VM\tablenotemark{b}}
}
\startdata
$\xi_t$ (km s$^{-1}$)  &1.85 &2.02 &1.9 \\
log A(C)               &8.50 &8.47 &8.67 \\
log A(O)               &8.87 &8.77 &8.75 \\
log A(Si)              &7.64 &7.67 &7.58 \\
log A(S)               &7.11 &7.08 &     \\
log A(Ca)              &6.43 &6.45 &6.31 \\
log A(Mn)              &5.29 &5.33 &     \\
log A(Fe)              &7.56 &7.58 &7.48 \\
log A(Ni)              &6.27 &6.33 &6.24 \\
log A(Ba)              &2.36 &2.30 &2.16
\enddata
\tablenotetext{a}{Stellar model temperature}
\tablenotetext{b}{Values from \citet{var99} with a 6696 K model}
\end{deluxetable}

\begin{deluxetable}{clccccccc}
\tablecaption{Adopted Stellar Parameters}
\tablewidth{0pt}
\tabletypesize{\footnotesize}
\tablenum{4}
\tablehead{
  \colhead{Star} & \colhead{Name} & \colhead{HD} & \colhead{V\tablenotemark{a}}
  & \colhead{B-V\tablenotemark{a}} & \colhead{MK Type\tablenotemark{a}}
  & \colhead{Gray\tablenotemark{b}} & \colhead{T$_{\rm eff}$\tablenotemark{c}} & \colhead{log $g$\tablenotemark{c}} \\
    \colhead{Number} & & & & & & &\colhead{(K)} &\colhead{(log cm s$^{-2}$)}
}
\startdata
\phn1 & $\beta$ Cas &\phn\phn\phn432 &2.28 &0.380 &F2 III-IV &F2 III  &6877 &3.43 \\
\phn2 & 9 Aur             &\phn32537 &4.98 &0.343 &F0 V      &F2 V    &7023 &4.07 \\
\phn3 & $\eta$ Lep        &\phn40136 &3.71 &0.337 &F1 III    &F1 V    &7117 &4.21 \\
\phn4 & $\rho$ Gem        &\phn58946 &4.16 &0.320 &F0 V      &F1 V    &6943 &4.05 \\
\phn5 & $\alpha$ CMi      &\phn61421 &0.40 &0.432 &F5 IV-V   &F5 IV-V &6618 &4.05 \\
\phn6 & $\chi$ Leo        &\phn96097 &4.62 &0.332 &F2 III-IV &F2 III  &7086 &3.91 \\
\phn7 & $\alpha$ Crv         &105452 &4.02 &0.334 &F2 III-IV &F0 IV-V &6961 &4.21 \\
\phn8 & 18 Boo               &125451 &5.41 &0.385 &F5 IV     &F3 V    &6739 &4.37 \\
\phn9 & $\theta$ Boo         &126660 &4.04 &0.497 &F7 V      &F7 V    &6371 &4.29 \\
10& $\sigma$ Boo             &128167 &4.47 &0.364 &F2 V      &F4 V    &6739 &4.37 \\
11& $\mu$ Vir                &129502 &3.87 &0.385 &F2 III    &F2 V    &6805 &4.24 \\
12& $\mu^{1}$ Boo            &137391 &4.31 &0.309 &F2 IVa    &F0 IV   &7253 &4.00 \\
13& $\sigma$ Ser             &147449 &4.82 &0.338 &F0 V      &F1 IV-V &7019 &4.03 \\
14& HR 6237                  &151613 &4.84 &0.375 &F2 V      &F2 V    &6722 &4.16 \\
15& $\xi$ Oph                &156987 &4.39 &0.394 &F1 III-IV &F2 V    &6723 &4.24
\enddata
\tablenotetext{a}{Values taken from the Hipparcos Catalog \citep{esa97}.}
\tablenotetext{b}{Uniform spectral types from \citet{gra89},
\citet{gra01}, and \citet{gra03}.}
\tablenotetext{c}{Calculated from Str\"{o}mgren photometry as described in \S\ 3.2.}
\end{deluxetable}

\begin{deluxetable}{clccccccccccc}
\tablecaption{Synthetic spectral fit results}
\tabletypesize{\scriptsize}
\tablewidth{0pt}
\tablenum{5}
\tablehead{
  \colhead{Star} & \colhead{Name} & \colhead{$\xi_{t}$}
  & \colhead{$v$ sin $i$} & \colhead{R/R$_\sun$\tablenotemark{a}}
  & \colhead{$P$/sin $i$\tablenotemark{a}}
  & \colhead{[Si/H]} & \colhead{[S/H]} & \colhead{[Ca/H]}
  & \colhead{[Ni/H]} & \colhead{[Fe/H]} & \colhead{[Fe/H]$_{lit}$}
  & \colhead{Ref.}\\
   & & \colhead{(km s$^{-1}$)}
   & \colhead{(km s$^{-1}$)} & & \colhead{(days)} & & & & &
}
\startdata
\phn1 &$\beta$ Cas   &3.8 &70.1    &3.43 &\phn2.48 &$-$0.25 &$-$0.13   &$-$0.15 &$-$0.42  &$-$0.14   &  &\\
\phn2 &9 Aur         &1.9 &21.0    &1.56 &\phn3.76 &$-$0.22 &$-$0.29   &$-$0.02 &$-$0.26  &$-$0.12   &$-$0.20 & 1\tablenotemark{b} \\
\phn3 &$\eta$ Lep    &2.2 &17.4    &1.62 &\phn4.71 &$-$0.05 &$-$0.12  &\phs0.10 &$-$0.10  &\phs0.00  &$-$0.05 & 1\tablenotemark{b} \\
\phn4 &$\rho$ Gem    &2.3 &58.3    &1.60 &\phn1.39 &$-$0.40 &$-$0.32   &$-$0.13 &$-$0.43  &$-$0.27   &  & \\
\phn5 &$\alpha$ CMi  &1.9 &\phn5.3 &1.94 &18.52   &\phs0.03 &$-$0.10  &\phs0.06 &$-$0.04  &\phs0.02  &$-$0.02 & 2\tablenotemark{b} \\
\phn6 &$\chi$ Leo    &3.4 &27.5    &1.99 &\phn3.66 &$-$0.12 &$-$0.11  &\phs0.22 &\phs0.05 &\phs0.04  &  & \\
\phn7 &$\alpha$ Crv  &1.8 &27.3    &1.36 &\phn2.52 &$-$0.20 &$-$0.16   &$-$0.06 &$-$0.30  &$-$0.12   &  & \\
\phn8 &18 Boo        &1.7 &38.9    &1.42 &\phn1.85 &$-$0.02 &\phs0.00 &\phs0.01 &$-$0.09  &$-$0.02   &$-$0.02 & 3 \\
\phn9 &$\theta$ Boo  &1.1 &31.8    &1.76 &\phn2.81 &$-$0.02 &$-$0.03  &\phs0.01 &$-$0.07  &$-$0.02   &$-$0.05 & 4 \\
   10 &$\sigma$ Boo  &1.5 &\phn9.0 &1.26 &\phn7.09 &$-$0.34 &$-$0.44   &$-$0.26 &$-$0.47  &$-$0.32   &$-$0.41 & 5\tablenotemark{b} \\
   11 &$\mu$ Vir     &2.0 &45.8    &1.99 &\phn2.20 &$-$0.21 &$-$0.18   &$-$0.13 &$-$0.27  &$-$0.17   & & \\
   12 &$\mu^{1}$ Boo &3.3 &82.3    &2.93 &\phn1.80 &$-$0.15 &\phs0.07  &$-$0.04 &$-$0.42  &$-$0.03   & & \\
   13 &$\sigma$ Ser  &3.0 &75.8    &1.75 &\phn1.16 &$-$0.15 &\phs0.12  &$-$0.13 &$-$0.20  &$-$0.04   & & \\
   14 &HR 6237       &2.0 &47.3    &1.86 &\phn1.99 &$-$0.34 &$-$0.18   &$-$0.21 &$-$0.43  &$-$0.22   & & \\
   15 &$\xi$ Oph     &1.7 &20.5    &1.45 &\phn3.57 &$-$0.22 &$-$0.21   &$-$0.10 &$-$0.31  &$-$0.17   &$-$0.13 & 6
\enddata
\tablerefs{(1) \citet{bur91}; (2) \citet{var99}; (3) \citet{boe88};
(4) \citet{bal90}; (5) \citet{edv93}; (6) \citet{edv84} }
\tablenotetext{a}{Not an observed quantity; calculated as described in \S\ 3.2}
\tablenotetext{b}{Other similar measurements are referenced in the \citet{cay97} catalog of [Fe/H] values}
\end{deluxetable}

\begin{deluxetable}{cccccccc}
\tablecaption{D$_3$ observations for $\rho$ Gem}
\tablewidth{0pt}
\tabletypesize{\footnotesize}
\tablenum{6}
\tablehead{
  \colhead{N} & \colhead{Run\tablenotemark{a}} & \colhead{JD--2400000} &
  \colhead{S/N} & \colhead{RV} & \colhead{$W_{\lambda}$} &
  \colhead {$\lambda_0$} & \colhead{FWHM} \\
  & & & \colhead{(pixel$^{-1}$)} & \colhead{(km s$^{-1}$)}
  & \colhead{(m\AA )} & \colhead{(\AA )} & \colhead{(km s$^{-1}$)}
}
\startdata
 1 &sp95 &49754.5677 &   170 & $-$5.12 &21.7 $\pm$ 3.0 &5875.760 $\pm$ 0.054 & \phn69 $\pm$ \phn7 \\
 2 &sp96 &50137.8234 &   130 & $-$5.94 &24.9 $\pm$ 5.0 &5875.715 $\pm$ 0.082 & \phn83 $\pm$ 13 \\
 3 &sp97 &50514.6102 &   191 & $-$5.43 &32.6 $\pm$ 3.7 &5875.819 $\pm$ 0.047 & \phn88 $\pm$ \phn8 \\
 4 &sp97 &50514.8189 &   167 & $-$4.76 &33.8 $\pm$ 4.1 &5875.779 $\pm$ 0.051 & \phn86 $\pm$ \phn8 \\
 5 &sp97 &50515.6139 &   113 & $-$5.04 &32.6 $\pm$ 6.3 &5875.671 $\pm$ 0.084 & \phn91 $\pm$ 14 \\
 6 &sp97 &50515.7910 &   279 & $-$4.10 &28.7 $\pm$ 2.2 &5875.717 $\pm$ 0.031 & \phn78 $\pm$ \phn5 \\
 7 &sp97 &50516.5837 &   186 & $-$5.52 &30.0 $\pm$ 3.3 &5875.845 $\pm$ 0.044 & \phn79 $\pm$ \phn7 \\
 8 &sp97 &50516.7915 &   192 & $-$4.49 &33.7 $\pm$ 3.9 &5875.729 $\pm$ 0.053 & \phn96 $\pm$ \phn9 \\
 9 &sp97 &50517.5834 &   195 & $-$5.52 &30.0 $\pm$ 3.3 &5875.898 $\pm$ 0.046 & \phn83 $\pm$ \phn7 \\
10 &sp97 &50517.7897 &   183 & $-$5.17 &22.8 $\pm$ 2.8 &5875.801 $\pm$ 0.043 & \phn64 $\pm$ \phn6 \\
11 &sp97 &50518.5817 &   155 & $-$5.05 &32.8 $\pm$ 3.9 &5875.738 $\pm$ 0.047 & \phn76 $\pm$ \phn7 \\
12 &sp97 &50518.7844 &   157 & $-$5.40 &34.5 $\pm$ 4.4 &5875.780 $\pm$ 0.056 & \phn89 $\pm$ \phn9 \\
13 &sp97 &50519.5822 &   162 & $-$4.15 &40.1 $\pm$ 4.2 &5875.834 $\pm$ 0.058 &    104 $\pm$ \phn9 \\
14 &sp97 &50519.7998 &   182 & $-$4.51 &41.0 $\pm$ 4.3 &5875.673 $\pm$ 0.046 & \phn97 $\pm$ \phn8 \\
15 &sp97 &50520.5844 &   196 & $-$4.62 &31.3 $\pm$ 2.9 &5875.795 $\pm$ 0.043 & \phn83 $\pm$ \phn6 \\
16 &sp97 &50520.7993 &   191 & $-$4.96 &35.3 $\pm$ 4.5 &5875.746 $\pm$ 0.057 &    104 $\pm$ 10 \\
17 &sp97 &50521.5959 &   221 & $-$5.07 &33.4 $\pm$ 3.2 &5875.841 $\pm$ 0.043 & \phn93 $\pm$ \phn7 \\
18 &sp97 &50521.8052 &   147 & $-$4.56 &29.8 $\pm$ 4.9 &5875.665 $\pm$ 0.075 & \phn95 $\pm$ 12 \\
19 &sp98 &50951.6176 &   173 & $-$3.68 &34.1 $\pm$ 3.5 &5875.652 $\pm$ 0.051 & \phn90 $\pm$ \phn7 \\
20 &fa99 &51562.6578 &\phn98 & $-$4.34 &23.0 $\pm$ 4.9 &5875.695 $\pm$ 0.071 & \phn60 $\pm$ 10
\enddata
\tablenotetext{a}{See Table 1 for observing run information.}
\end{deluxetable}

\begin{deluxetable}{cccccccc}
\tablecaption{D$_3$ observations for $\theta$ Boo}
\tablewidth{0pt}
\tabletypesize{\footnotesize}
\tablenum{7}
\tablehead{
  \colhead{Star} & \colhead{Run\tablenotemark{a}} & \colhead{JD--2400000} &
  \colhead{S/N} & \colhead{RV} & \colhead{$W_{\lambda}$} &
  \colhead {$\lambda_0$} & \colhead{FWHM} \\
  & & & \colhead{(pixel$^{-1}$)} & \colhead{(km s$^{-1}$)}
  & \colhead{(m\AA )} & \colhead{(\AA )} & \colhead{(km s$^{-1}$)}
}
\startdata
 1 &sp95 &49751.0282 &   189 & $-$\phn9.96 &37.1 $\pm$ 2.3 &5875.777 $\pm$ 0.019 & 52.2 $\pm$ 2.5 \\
 2 &sp95 &49751.0366 &   181 &    $-$10.24 &40.5 $\pm$ 2.8 &5875.671 $\pm$ 0.019 & 56.3 $\pm$ 3.0 \\
 3 &sp97 &50518.0140 &   196 & $-$\phn9.86 &42.7 $\pm$ 2.7 &5875.713 $\pm$ 0.017 & 55.1 $\pm$ 2.6 \\
 4 &sp97 &50518.8916 &   176 &    $-$10.03 &41.8 $\pm$ 2.5 &5875.764 $\pm$ 0.019 & 54.0 $\pm$ 2.5 \\
 5 &sp97 &50519.0280 &   185 &    $-$10.45 &40.8 $\pm$ 2.6 &5875.757 $\pm$ 0.018 & 54.2 $\pm$ 2.6 \\
 6 &sp97 &50519.8902 &   201 &    $-$10.25 &40.7 $\pm$ 2.7 &5875.661 $\pm$ 0.018 & 55.8 $\pm$ 2.9 \\
 7 &sp97 &50521.0371 &   209 &    $-$10.19 &41.0 $\pm$ 2.4 &5875.672 $\pm$ 0.015 & 51.8 $\pm$ 2.4 \\
 8 &sp97 &50521.8947 &   215 &    $-$10.48 &38.1 $\pm$ 2.4 &5875.719 $\pm$ 0.016 & 52.8 $\pm$ 2.5 \\
 9 &sp97 &50522.0147 &   237 &    $-$10.61 &42.8 $\pm$ 2.3 &5875.721 $\pm$ 0.015 & 56.9 $\pm$ 2.4 \\
10 &sp98 &50945.6186 &   227 &    $-$10.19 &34.5 $\pm$ 2.2 &5875.722 $\pm$ 0.017 & 51.8 $\pm$ 2.5 \\
11 &sp98 &50945.9428 &   196 &    $-$10.34 &35.4 $\pm$ 2.2 &5875.702 $\pm$ 0.018 & 51.5 $\pm$ 2.4 \\
12 &sp98 &50946.6136 &   168 &    $-$10.32 &33.0 $\pm$ 2.9 &5875.708 $\pm$ 0.024 & 52.4 $\pm$ 3.7 \\
13 &sp98 &50946.8060 &   116 &    $-$10.37 &35.3 $\pm$ 4.3 &5875.630 $\pm$ 0.032 & 52.4 $\pm$ 5.1 \\
14 &sp98 &50948.6099 &   203 &    $-$10.94 &32.5 $\pm$ 2.1 &5875.756 $\pm$ 0.021 & 53.4 $\pm$ 2.8 \\
15 &sp98 &50948.7982 &   244 & $-$\phn9.90 &37.8 $\pm$ 2.3 &5875.690 $\pm$ 0.017 & 58.7 $\pm$ 2.9 \\
16 &sp98 &50948.9714 &   222 & $-$\phn9.93 &34.1 $\pm$ 2.4 &5875.741 $\pm$ 0.019 & 55.4 $\pm$ 3.0 \\
17 &sp98 &50949.6164 &   224 &    $-$10.64 &27.0 $\pm$ 1.8 &5875.707 $\pm$ 0.019 & 48.2 $\pm$ 2.4 \\
18 &sp98 &50949.8087 &   229 &    $-$10.64 &36.5 $\pm$ 2.3 &5875.685 $\pm$ 0.017 & 55.6 $\pm$ 2.8 \\
19 &sp98 &50949.9499 &\phn91 &    $-$11.43 &35.1 $\pm$ 4.6 &5875.705 $\pm$ 0.039 & 50.5 $\pm$ 5.2 \\
20 &sp98 &50950.6263 &   134 &    $-$10.27 &32.4 $\pm$ 3.0 &5875.670 $\pm$ 0.027 & 47.9 $\pm$ 3.5 \\
21 &sp98 &50951.9703 &   228 &    $-$10.02 &32.5 $\pm$ 1.8 &5875.659 $\pm$ 0.017 & 50.6 $\pm$ 2.3 \\
22 &fa99 &51563.0687 &   163 & $-$\phn9.71 &37.2 $\pm$ 3.2 &5875.691 $\pm$ 0.023 & 55.3 $\pm$ 3.7
\enddata
\tablenotetext{a}{See Table 1 for observing run information.}
\end{deluxetable}

\begin{deluxetable}{cccccccc}
\tablecaption{D$_3$ observations for $\mu$ Vir}
\tablewidth{0pt}
\tabletypesize{\footnotesize}
\tablenum{8}
\tablehead{
  \colhead{N} & \colhead{Run\tablenotemark{a}} & \colhead{JD--2400000} &
  \colhead{S/N} & \colhead{RV} & \colhead{$W_{\lambda}$} &
  \colhead {$\lambda_0$} & \colhead{FWHM} \\
  & & & \colhead{(pixel$^{-1}$)} & \colhead{(km s$^{-1}$)} &
  \colhead{(m\AA )} & \colhead{(\AA )} & \colhead{(km s$^{-1}$)}
}
\startdata
 1 &fa95 &50056.0422 &174 &4.57 &28.9 $\pm$ 2.6 &5875.694 $\pm$ 0.029 &55 $\pm$ \phn4 \\
 2 &sp97 &50519.8754 &216 &3.37 &28.8 $\pm$ 2.2 &5875.785 $\pm$ 0.029 &64 $\pm$ \phn4 \\
 3 &sp98 &50945.6358 &201 &2.57 &27.1 $\pm$ 2.8 &5875.885 $\pm$ 0.041 &74 $\pm$ \phn6 \\
 4 &sp98 &50945.6910 &181 &2.46 &25.4 $\pm$ 3.0 &5875.759 $\pm$ 0.046 &71 $\pm$ \phn7 \\
 5 &sp98 &50945.7477 &198 &2.84 &21.9 $\pm$ 3.0 &5875.740 $\pm$ 0.057 &77 $\pm$ \phn8 \\
 6 &sp98 &50945.9259 &163 &2.68 &22.5 $\pm$ 3.1 &5875.718 $\pm$ 0.049 &64 $\pm$ \phn7 \\
 7 &sp98 &50946.6285 &150 &3.00 &33.2 $\pm$ 4.1 &5875.613 $\pm$ 0.054 &83 $\pm$ \phn8 \\
 8 &sp98 &50946.6822 &146 &2.55 &27.7 $\pm$ 4.3 &5875.823 $\pm$ 0.068 &85 $\pm$ 11 \\
 9 &sp98 &50946.7441 &164 &3.06 &28.1 $\pm$ 3.3 &5875.843 $\pm$ 0.045 &71 $\pm$ \phn7 \\
10 &sp98 &50947.9247 &213 &2.71 &28.6 $\pm$ 2.8 &5875.608 $\pm$ 0.044 &85 $\pm$ \phn7 \\
11 &sp98 &50948.6318 &236 &2.57 &29.3 $\pm$ 2.3 &5875.727 $\pm$ 0.030 &70 $\pm$ \phn4 \\
12 &sp98 &50948.6841 &231 &2.54 &31.6 $\pm$ 2.8 &5875.896 $\pm$ 0.039 &87 $\pm$ \phn6 \\
13 &sp98 &50948.7387 &245 &3.08 &30.2 $\pm$ 2.3 &5875.773 $\pm$ 0.031 &76 $\pm$ \phn5 \\
14 &sp98 &50948.8129 &261 &3.28 &36.5 $\pm$ 2.3 &5875.777 $\pm$ 0.028 &80 $\pm$ \phn4 \\
15 &sp98 &50948.8647 &262 &3.32 &35.0 $\pm$ 2.3 &5875.749 $\pm$ 0.028 &81 $\pm$ \phn4 \\
16 &sp98 &50948.9192 &248 &3.14 &28.8 $\pm$ 2.3 &5875.732 $\pm$ 0.033 &77 $\pm$ \phn5 \\
17 &sp98 &50949.6310 &231 &2.15 &25.7 $\pm$ 2.3 &5875.802 $\pm$ 0.034 &69 $\pm$ \phn5 \\
18 &sp98 &50949.6832 &252 &2.68 &24.1 $\pm$ 2.1 &5875.796 $\pm$ 0.032 &67 $\pm$ \phn5 \\
19 &sp98 &50949.7380 &267 &2.66 &29.6 $\pm$ 2.4 &5875.780 $\pm$ 0.035 &85 $\pm$ \phn5 \\
20 &sp98 &50949.8216 &260 &2.98 &30.2 $\pm$ 2.4 &5875.751 $\pm$ 0.034 &83 $\pm$ \phn5 \\
21 &sp98 &50949.8701 &227 &3.27 &32.8 $\pm$ 2.5 &5875.673 $\pm$ 0.036 &84 $\pm$ \phn5 \\
22 &sp98 &50949.9214 &170 &2.97 &28.5 $\pm$ 3.8 &5875.697 $\pm$ 0.060 &85 $\pm$ \phn9 \\
23 &sp98 &50950.7731 &217 &2.70 &23.9 $\pm$ 2.5 &5875.656 $\pm$ 0.042 &71 $\pm$ \phn6 \\
24 &sp98 &50951.6531 &275 &2.36 &24.9 $\pm$ 2.0 &5875.608 $\pm$ 0.030 &70 $\pm$ \phn4 \\
25 &sp98 &50951.7112 &260 &2.72 &23.7 $\pm$ 2.0 &5875.695 $\pm$ 0.032 &67 $\pm$ \phn4 \\
26 &sp98 &50951.7644 &264 &2.67 &31.0 $\pm$ 2.4 &5875.755 $\pm$ 0.034 &85 $\pm$ \phn5 \\
27 &sp98 &50951.8680 &265 &3.21 &30.6 $\pm$ 2.4 &5875.615 $\pm$ 0.033 &83 $\pm$ \phn5 \\
28 &sp98 &50951.9204 &251 &2.75 &29.9 $\pm$ 2.5 &5875.655 $\pm$ 0.035 &83 $\pm$ \phn5 \\
29 &fa99 &51570.9607 &202 &3.86 &24.4 $\pm$ 2.4 &5875.801 $\pm$ 0.034 &62 $\pm$ \phn5
\enddata
\tablenotetext{a}{See Table 1 for observing run information.}
\end{deluxetable}

\begin{deluxetable}{cccccccc}
\tablecaption{D$_3$ observations for 18 Boo}
\tablewidth{0pt}
\tabletypesize{\footnotesize}
\tablenum{9}
\tablehead{
  \colhead{N} & \colhead{Run\tablenotemark{a}} & \colhead{JD--2400000} &
  \colhead{S/N} & \colhead{RV} & \colhead{$W_{\lambda}$} &
  \colhead {$\lambda_0$} & \colhead{FWHM} \\
  & & & \colhead{(pixel$^{-1}$)} & \colhead{(km s$^{-1}$)}
  & \colhead{(m\AA )} & \colhead{(\AA )} & \colhead{(km s$^{-1}$)}
}
\startdata
 1 &sp95 &49751.0467 &    124 & \phs0.21 &37.0 $\pm$ 5.0 & 5875.635$\pm$ 0.043 &69 $\pm$ \phn7 \\
 2 &sp96 &50137.0501 & \phn80 & \phs0.46 &30.1 $\pm$ 5.8 & 5875.608$\pm$ 0.065 &56 $\pm$ \phn9 \\
 3 &sp97 &50515.0275 &    126 & \phs0.26 &23.3 $\pm$ 4.3 & 5875.752$\pm$ 0.051 &55 $\pm$ \phn8 \\
 4 &sp97 &50516.0100 & \phn66 & \phs0.51 &41.1 $\pm$ 9.9 & 5875.448$\pm$ 0.083 &73 $\pm$ 13 \\
 5 &sp98 &50945.6632 &    167 &  $-$0.37 &20.2 $\pm$ 2.7 & 5875.784$\pm$ 0.049 &59 $\pm$ \phn6 \\
 6 &sp98 &50945.7209 &    164 & \phs0.33 &28.4 $\pm$ 3.1 & 5875.553$\pm$ 0.044 &70 $\pm$ \phn6 \\
 7 &sp98 &50945.8991 &    157 & \phs0.84 &28.3 $\pm$ 3.0 & 5875.675$\pm$ 0.041 &65 $\pm$ \phn6 \\
 8 &sp98 &50946.6551 &    140 &  $-$0.38 &25.6 $\pm$ 3.7 & 5875.613$\pm$ 0.059 &73 $\pm$ \phn8 \\
 9 &sp98 &50946.7134 &    145 &  $-$0.33 &21.0 $\pm$ 2.7 & 5875.753$\pm$ 0.038 &49 $\pm$ \phn5 \\
10 &sp98 &50946.7745 &    147 &  $-$0.15 &19.1 $\pm$ 3.3 & 5875.691$\pm$ 0.064 &66 $\pm$ \phn9 \\
11 &sp98 &50947.8999 &    186 & \phs0.34 &28.8 $\pm$ 2.7 & 5875.643$\pm$ 0.036 &67 $\pm$ \phn5 \\
12 &sp98 &50948.6580 &    206 & \phs0.00 &28.1 $\pm$ 2.4 & 5875.700$\pm$ 0.033 &67 $\pm$ \phn4 \\
13 &sp98 &50948.7127 &    197 &  $-$0.06 &20.6 $\pm$ 2.2 & 5875.659$\pm$ 0.036 &55 $\pm$ \phn5 \\
14 &sp98 &50948.7653 &    215 & \phs0.25 &25.8 $\pm$ 2.3 & 5875.716$\pm$ 0.034 &67 $\pm$ \phn5 \\
15 &sp98 &50948.8388 &    214 & \phs0.43 &25.4 $\pm$ 2.1 & 5875.666$\pm$ 0.030 &60 $\pm$ \phn4 \\
16 &sp98 &50948.8933 &    212 & \phs0.65 &32.4 $\pm$ 2.5 & 5875.725$\pm$ 0.032 &73 $\pm$ \phn4 \\
17 &sp98 &50948.9453 &    195 & \phs0.23 &32.0 $\pm$ 2.8 & 5875.702$\pm$ 0.039 &78 $\pm$ \phn5 \\
18 &sp98 &50949.6572 &    197 &  $-$0.18 &19.4 $\pm$ 2.2 & 5875.728$\pm$ 0.036 &54 $\pm$ \phn5 \\
19 &sp98 &50949.7119 &    217 & \phs0.36 &20.6 $\pm$ 2.1 & 5875.718$\pm$ 0.035 &58 $\pm$ \phn5 \\
20 &sp98 &50949.7628 &    213 &  $-$0.34 &18.0 $\pm$ 2.0 & 5875.647$\pm$ 0.035 &52 $\pm$ \phn5 \\
21 &sp98 &50949.8459 &    216 & \phs0.68 &21.3 $\pm$ 2.0 & 5875.641$\pm$ 0.032 &56 $\pm$ \phn4 \\
22 &sp98 &50949.8970 &    192 & \phs0.23 &22.5 $\pm$ 2.5 & 5875.669$\pm$ 0.041 &64 $\pm$ \phn6 \\
23 &sp98 &50950.7430 &    172 &  $-$0.59 &24.0 $\pm$ 2.9 & 5875.679$\pm$ 0.047 &69 $\pm$ \phn6 \\
24 &sp98 &50950.8026 &    157 & \phs0.25 &26.0 $\pm$ 3.0 & 5875.697$\pm$ 0.044 &63 $\pm$ \phn6 \\
25 &sp98 &50951.6818 &    185 &  $-$0.44 &15.5 $\pm$ 2.1 & 5875.702$\pm$ 0.036 &45 $\pm$ \phn5 \\
26 &sp98 &50951.7378 &    202 &  $-$0.29 &20.1 $\pm$ 2.3 & 5875.659$\pm$ 0.042 &60 $\pm$ \phn5 \\
27 &sp98 &50951.8202 &    160 &  $-$1.22 &29.2 $\pm$ 3.2 & 5875.663$\pm$ 0.043 &69 $\pm$ \phn5 \\
28 &sp98 &50951.8956 &    215 & \phs0.23 &24.8 $\pm$ 2.3 & 5875.620$\pm$ 0.036 &68 $\pm$ \phn5 \\
29 &sp98 &50951.9453 &    191 &  $-$0.53 &20.7 $\pm$ 2.2 & 5875.600$\pm$ 0.035 &55 $\pm$ \phn4
\enddata
\tablenotetext{a}{See Table 1 for observing run information.}
\end{deluxetable}

\begin{deluxetable}{cccccccccccc}
\tablecaption{Short-term variability results}
\tablewidth{0pt}
\tabletypesize{\footnotesize}
\tablenum{10}
\tablehead{
  \colhead{Star} & \colhead{Run\tablenotemark{a}} & \colhead{$N_{\rm int}$} &
  \colhead{$\langle W_{\lambda} \rangle$}
  & \colhead{SD\tablenotemark{b}}
  & \colhead{$\langle$Err$\rangle$\tablenotemark{c}}
  & \colhead{$\langle \lambda_0 \rangle$}
  & \colhead{SD\tablenotemark{b}}
  & \colhead{$\langle$Err$\rangle$\tablenotemark{c}}
  & \colhead{$\langle$FWHM$\rangle$}
  & \colhead{SD\tablenotemark{b}}
  & \colhead{$\langle$Err$\rangle$\tablenotemark{c}} \\
  & & & \colhead{(m\AA )} & \colhead{(m\AA )} & \colhead{(m\AA )} &
  \colhead{(\AA )} & \colhead{(\AA )}  & \colhead{(\AA )} &
  \colhead{(km s$^{-1}$)} & \colhead{(km s$^{-1}$)} & \colhead{(km s$^{-1}$)}
}
\startdata
$\rho$ Gem   &sp97 &    16 &31.5 &4.4 &3.9 &5875.777 &0.066 &0.052 &84.6    &10.7 & 8.2 \\
$\theta$ Boo &sp97 & \phn7 &41.1 &1.6 &2.5 &5875.713 &0.037 &0.017 &54.3 &\phn1.8 & 2.6 \\
$\theta$ Boo &sp98 &    12 &33.3 &3.2 &2.7 &5875.701 &0.031 &0.022 &52.3 &\phn3.1 & 3.2 \\
$\mu$ Vir    &sp98 &    26 &28.5 &3.9 &2.7 &5875.735 &0.077 &0.040 &76.6 &\phn7.0 & 5.8 \\
18 Boo       &sp98 &    25 &23.2 &4.5 &2.6 &5875.677 &0.048 &0.040 &61.3 &\phn8.2 & 5.3
\enddata
\tablenotetext{a}{See Table 1 for observing run information.}
\tablenotetext{b}{Weighted standard deviation of the measurements.}
\tablenotetext{c}{Weighted mean of the measurement uncertainties.}

\end{deluxetable}

\begin{deluxetable}{ccccccccc}
\tablecaption{Variability probabilities using the $\chi^2$ test}
\tablewidth{0pt}
\tabletypesize{\footnotesize}
\tablenum{11}
\tablehead{
  & & & \multicolumn{3}{c}{Reported errors} & \multicolumn{3}{c}{15\% greater errors} \\
  \cline{4-6} \cline{7-9} \\
  \colhead{Star} & \colhead{Run\tablenotemark{a}} & \colhead{$N_{\rm int}$} &
  \colhead{$P_{\rm W\lambda}$} & \colhead{$P_{\rm \lambda 0}$} & \colhead{$P_{\rm FWHM}$} &
  \colhead{$P_{\rm W\lambda}$} & \colhead{$P_{\rm \lambda 0}$} & \colhead{$P_{\rm FWHM}$} 
}
\startdata
$\rho$ Gem   &sp97 &    16 &\phm{$>$}0.922 &\phm{$>$}0.980 &\phm{$>$}0.991  &\phm{$>$}0.716 &\phm{$>$}0.875 &\phm{$>$}0.924 \\
$\theta$ Boo &sp97 & \phn7 &\phm{$>$}0.140       &$>$0.999 &\phm{$>$}0.183  &\phm{$>$}0.076 &\phm{$>$}0.999 &\phm{$>$}0.101 \\
$\theta$ Boo &sp98 &    12 &\phm{$>$}0.956 &\phm{$>$}0.995 &\phm{$>$}0.696  &\phm{$>$}0.828 &\phm{$>$}0.957 &\phm{$>$}0.443 \\
$\mu$ Vir    &sp98 &    26 &$>$0.999             &$>$0.999 &\phm{$>$}0.983  &\phm{$>$}0.987       &$>$0.999 &\phm{$>$}0.838 \\
18 Boo       &sp98 &    25 &$>$0.999       &\phm{$>$}0.960       &$>$0.999        &$>$0.999 &\phm{$>$}0.752 &\phm{$>$}0.998
\enddata
\tablenotetext{a}{See Table 1 for observing run information.}
\end{deluxetable}

\begin{deluxetable}{cccccccccccc}
\tablecaption{Photospheric line fits for 18 Boo during sp98}
\tablewidth{0pt}
\tabletypesize{\footnotesize}
\tablenum{12}
\tablehead{
  \colhead{Line} & \colhead{$N_{\rm int}$} &
  \colhead{$\langle W_{\lambda} \rangle$}
  & \colhead{SD\tablenotemark{a}}
  & \colhead{$\langle$Err$\rangle$\tablenotemark{b}}
  & \colhead{$\langle \lambda_0 \rangle$}
  & \colhead{SD\tablenotemark{a}}
  & \colhead{$\langle$Err$\rangle$\tablenotemark{b}}
  & \colhead{$\langle$FWHM$\rangle$}
  & \colhead{SD\tablenotemark{a}}
  & \colhead{$\langle$Err$\rangle$\tablenotemark{b}} \\
  & & \colhead{(m\AA )} & \colhead{(m\AA )} & \colhead{(m\AA )} &
  \colhead{(\AA )} & \colhead{(\AA )}  & \colhead{(\AA )} &
  \colhead{(km s$^{-1}$)} & \colhead{(km s$^{-1}$)} & \colhead{(km s$^{-1}$)}
}
\startdata
Fe I $\lambda$5816  & 25 &55.2 &3.5 &3.6 &5816.375 &0.014 & 0.016 & 54.2 & 2.6 & 2.8 \\
Fe I $\lambda$5848  & 25 &16.5 &2.4 &2.2 &5848.090 &0.043 & 0.038 & 54.1 & 5.9 & 5.8 \\
Fe I $\lambda$5934  & 25 &52.1 &2.6 &3.0 &5934.636 &0.017 & 0.017 & 55.3 & 2.3 & 2.6 \\

\enddata
\tablenotetext{a}{Weighted standard deviation of the measurements.}
\tablenotetext{b}{Weighted mean of the measurement uncertainties.}
\end{deluxetable}

\begin{deluxetable}{cccccccc}
\tablecaption{Additional D$_3$ observations}
\tablewidth{0pt}
\tabletypesize{\footnotesize}
\tablenum{13}
\tablehead{
  \colhead{Star} & \colhead{Run\tablenotemark{a}} & \colhead{JD--2400000} &
  \colhead{S/N} & \colhead{RV} & \colhead{$W_{\lambda}$} &
  \colhead {$\lambda_0$} & \colhead{FWHM} \\
  & & & \colhead{(pixel$^{-1}$)} & \colhead{(km s$^{-1}$)}
  & \colhead{(m\AA )} & \colhead{(\AA )} & \colhead{(km s$^{-1}$)}
}
\startdata
 $\beta$ Cas &fa95 &50052.5386 &   205 &\phn\phs5.94 &   21.8 $\pm$ 4.5 &5875.381 $\pm$ 0.138 &   142 $\pm$ 22\\
 $\beta$ Cas &sp98 &50948.9820 &   219 &\phs10.40    &   23.9 $\pm$ 3.9 &5875.408 $\pm$ 0.132 &   154 $\pm$ 20\\
 $\beta$ Cas &fa99 &51570.5511 &   311 &\phn\phs7.75 &   14.8 $\pm$ 2.2 &5875.453 $\pm$ 0.100 &   119 $\pm$ 14\\

       9 Aur &sp96 &50136.5929 &\phn67 &\phn$-$1.57 &    15.6 $\pm$ 6.2 &5875.996 $\pm$ 0.112 &\phn47 $\pm$ 15\\
       9 Aur &sp97 &50521.6403 &   152 &\phn$-$2.58 &    20.3 $\pm$ 2.7 &5875.823 $\pm$ 0.040 &\phn50 $\pm$ \phn5\\

  $\eta$ Lep &fa95 &50057.7213 &   165 &\phn$-$0.78 &    35.6 $\pm$ 2.7 &5875.699 $\pm$ 0.020 &\phn47 $\pm$ \phn2\\
  $\eta$ Lep &fa96 &50373.9985 &   215 &\phn$-$0.91 &    31.2 $\pm$ 2.1 &5875.778 $\pm$ 0.018 &\phn51 $\pm$ \phn2\\
  $\eta$ Lep &sp97 &50520.5941 &   199 &\phn$-$0.86 &    32.6 $\pm$ 2.2 &5875.746 $\pm$ 0.022 &\phn53 $\pm$ \phn3\\

$\alpha$ CMi &sp96 &50137.5669 &   183 &\phn$-$3.00  &\phn6.5 $\pm$ 2.5 &5875.693 $\pm$ 0.129 &\phn56 $\pm$ 17\\
$\alpha$ CMi &sp96 &50142.7192 &   281 &\phn$-$2.78  &\phn8.8 $\pm$ 1.6 &5875.451 $\pm$ 0.061 &\phn56 $\pm$ \phn8\\
$\alpha$ CMi &sp97 &50521.6544 &   314 &\phn$-$2.31  &\phn7.0 $\pm$ 1.5 &5875.707 $\pm$ 0.069 &\phn57 $\pm$ \phn9\\

  $\chi$ Leo &sp95 &49754.0362 &   263 &\phn\phs4.51 &\phn3.7 $\pm$ 1.2 &5875.556 $\pm$ 0.060 &\phn31 $\pm$ \phn8\\
  $\chi$ Leo &fa96 &50374.0236 &   245 &\phn\phs5.80 &\phn6.3 $\pm$ 1.9 &5875.602 $\pm$ 0.108 &\phn63 $\pm$ 15\\
  $\chi$ Leo &sp97 &50520.9079 &   214 &\phn\phs6.58 &\phn4.7 $\pm$ 2.0 &5875.587 $\pm$ 0.106 &\phn42 $\pm$ 15\\
  $\chi$ Leo &sp98 &50951.6350 &   162 &\phn\phs6.40 &  \ldots       & \ldots              & \ldots \\
  $\chi$ Leo &fa99 &51565.0614 &   158 &\phn\phs6.05 &\phn4.5 $\pm$ 1.9 &5875.577 $\pm$ 0.068 &\phn27 $\pm$ \phn9\\

$\alpha$ Crv &sp95 &49754.0651 &   226 &\phn\phs3.00 &   21.5 $\pm$ 1.8 &5875.749 $\pm$ 0.024 &\phn50 $\pm$ \phn3\\
$\alpha$ Crv &sp97 &50517.8913 &   187 &\phn\phs4.87 &   26.3 $\pm$ 2.5 &5875.760 $\pm$ 0.037 &\phn65 $\pm$ \phn5\\

$\sigma$ Boo &fa95 &50052.0561 &   187 &\phn\phs1.66 &\phn7.3 $\pm$ 1.6 &5875.695 $\pm$ 0.040 &\phn29 $\pm$ \phn5\\
$\sigma$ Boo &sp97 &50521.9554 &   145 &\phn\phs1.05 &   13.7 $\pm$ 2.1 &5875.640 $\pm$ 0.028 &\phn29 $\pm$ \phn3\\

 $\mu^1$ Boo &fa95 &50056.0338 &   167 &\phn$-$9.08 &    17.5 $\pm$ 3.8 &5875.567 $\pm$ 0.075 &\phn74 $\pm$ 13\\
 $\mu^1$ Boo &sp97 &50519.0146 &   177 &\phn$-$7.09 &    24.1 $\pm$ 7.0 &5875.610 $\pm$ 0.107 &   115 $\pm$ 24\\
 $\mu^1$ Boo &sp98 &50951.8521 &\phn91 &\phn$-$1.83 &   \ldots       & \ldots              & \ldots \\

$\sigma$ Ser &sp95 &49752.0628 &   189 &$-$49.33 &      35.0 $\pm$ 4.5 &5875.966 $\pm$ 0.057 &   107 $\pm$ 10\\
$\sigma$ Ser &sp97 &50521.9777 &   181 &$-$50.77 &      28.0 $\pm$ 4.1 &5875.891 $\pm$ 0.064 &\phn95 $\pm$ 11\\

     HR 6237 &fa95 &50050.5463 &   152 &$-$18.96 &      23.1 $\pm$ 2.9 &5875.873 $\pm$ 0.047 &\phn61 $\pm$ \phn6\\
     HR 6237 &sp97 &50521.9927 &   173 &\phn\phs0.64 &  23.6 $\pm$ 2.9 &5875.740 $\pm$ 0.040 &\phn60 $\pm$ \phn6\\

   $\xi$ Oph &sp96 &50143.0288 &   113 &\phn$-$9.51 &   19.4 $\pm$ 3.1 &5875.762 $\pm$ 0.039 &\phn39 $\pm$ \phn5\\
   $\xi$ Oph &sp97 &50522.0047 &   153 &\phn$-$8.25 &   21.9 $\pm$ 2.5 &5875.745 $\pm$ 0.032 &\phn47 $\pm$ \phn4\\
\enddata
\tablenotetext{a}{See Table 1 for observing run information.}
\end{deluxetable}

\end{document}